\pdfoutput=1

\documentclass{article} %
\usepackage{acl}

\usepackage{amsmath,amsfonts,bm}

\def\eqref#1{equation~\ref{#1}}

\def\1{\bm{1}}

\DeclareMathAlphabet{\mathsfit}{\encodingdefault}{\sfdefault}{m}{sl}
\SetMathAlphabet{\mathsfit}{bold}{\encodingdefault}{\sfdefault}{bx}{n}

\usepackage{hyperref}
\usepackage{url}
\usepackage{enumitem}
\usepackage[T1]{fontenc}
\usepackage{times}
\usepackage{latexsym}
\usepackage{graphicx}
\usepackage{booktabs}
\usepackage{multirow}
\usepackage{subcaption}
\usepackage{fancyvrb,fvextra}
\usepackage{wrapfig}
\usepackage{tikz}
\usepackage{arydshln}
\usepackage{pgfplots}
\pgfplotsset{compat=1.18}

\usepackage{floatrow}
\newfloatcommand{capbtabbox}{table}[][\FBwidth]

\usepackage{listings,xcolor}
\definecolor{codegray}{rgb}{0.95,0.95,0.95}
\lstnewenvironment{longprompt}[1][]
{\lstset{basicstyle={\small\it},prebreak={},postbreak=\mbox{\hspace{-2.25 em}},backgroundcolor=\color{codegray},frame=single,framerule=1pt,#1}}
{}
\usepackage{amssymb}%
\usepackage{pifont}%
\newcommand{\cmark}{\textcolor[RGB]{0,100,0}{\ding{51}}}%
\newcommand{\xmark}{\textcolor{red}{\ding{55}}}%

\title{Scaling Rich Style-Prompted Text-to-Speech Datasets}

\author{Anuj Diwan$^\heartsuit$, Zhisheng Zheng$^\heartsuit$, David Harwath$^\heartsuit$, Eunsol Choi$^\clubsuit$ \\
Department of Computer Science, The University of Texas at Austin$^\heartsuit$ \\
Department of Computer Science and Data Science, New York University$^\clubsuit$  \\
\texttt{\{anuj.diwan,zszheng,harwath\}@utexas.edu}, \texttt{eunsol@nyu.edu}}

\begin{document}

\maketitle

\begin{abstract}
We introduce \textbf{Para}linguistic \textbf{Speech} \textbf{Cap}tion\textbf{s} (\textbf{ParaSpeechCaps}), a large-scale dataset that annotates speech utterances with rich style captions. While rich abstract tags (e.g. \textit{guttural, nasal, pained}) have been explored in small-scale human-annotated datasets, existing large-scale datasets only cover basic tags (e.g. \textit{low-pitched, slow, loud}). We combine off-the-shelf text and speech embedders, classifiers and an audio language model to automatically scale rich tag annotations for the first time. ParaSpeechCaps covers a total of $59$ style tags, including both speaker-level intrinsic tags and utterance-level situational tags. It consists of $282$ hours of human-labelled data (PSC-Base) and $2427$ hours of automatically annotated data (PSC-Scaled). We finetune Parler-TTS, an open-source style-prompted TTS model, on ParaSpeechCaps, and achieve improved style consistency (+7.9\% Consistency MOS) and speech quality (+15.5\% Naturalness MOS) over the best performing baseline that combines existing rich style tag datasets. We ablate several of our dataset design choices to lay the foundation for future work in this space. Our dataset, models and code are released at~\url{https://github.com/ajd12342/paraspeechcaps}.
\end{abstract}

\section{Introduction}
\begin{figure*}
    \centering
    \includegraphics[width=\linewidth]{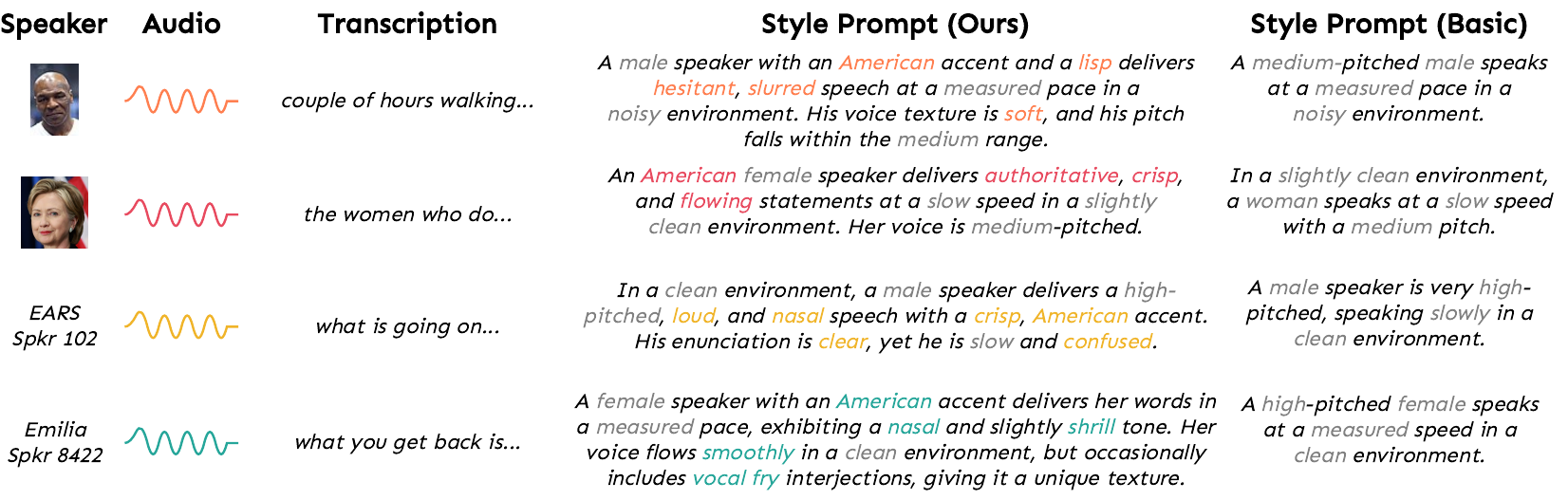}
    \caption{Randomly sampled examples from ParaSpeechCaps. Our style prompts cover rich tags describing complex styles like rhythm, clarity, emotion, etc. in contrast to erstwhile basic style prompts that only contain gender, pitch and speed levels. We highlight rich style tags with vibrant colors and basic style tags with a gray color.\vspace{-10pt}}
    \label{fig:datasetexamples}
\end{figure*}
Style-prompted text-to-speech models~\cite{guo2022promptttscontrollabletexttospeechtext,leng2023prompttts2describinggenerating,lacombe-etal-2024-parler-tts} can synthesize speech while controlling for style factors like pitch, speed and emotion via textual style prompts. Building such a system requires a training dataset where each example consists of a transcript, a style prompt and an utterance reflecting the specified style prompt. Yet, such data is often costly to annotate and existing datasets~\cite{kawamura2024librittspcorpusspeakingstyle,lacombe-etal-2024-parler-tts,Ji2024} are either limited in their scale or their coverage of style tag types.

In this paper, we introduce Paralinguistic Speech Captions (\textbf{ParaSpeechCaps}), a dataset which covers $59$ unique style tags.  We categorize style tags into intrinsic tags tied to a speaker's identity (e.g., \textit{shrill, guttural}) and situational tags that characterize individual utterances (e.g., \textit{happy, whispered}). Our dataset consists of a human-annotated portion (\textbf{PSC-Base}, $282$ hrs) and an automatically labeled portion (\textbf{PSC-Scaled}, $2427$ hrs), covering $33$ intrinsic and $26$ situational tags. Figure~\ref{fig:datasetexamples} shows a few examples. We first build PSC-Base by aggregating existing situational annotations as well as collecting new intrinsic annotations on $282$ hours of speech~\citep{nguyen2023expressobenchmarkanalysisdiscrete,richter2024earsanechoicfullbandspeech,nagrani2020voxceleb} via crowdsourcing.

As the human-annotated dataset is limited in scale, we propose two novel data scaling approaches to expand it, one for intrinsic tags and one for situational tags (Figure~\ref{fig:scaledfig}). We source speech and transcripts from the $45$k-hr English portion of a large-scale speaker-labeled corpus~\cite{he2024emiliaextensivemultilingualdiverse} and apply both approaches to identify instances with the target style tag. Existing large-scale datasets~\citep{lacombe-etal-2024-parler-tts,lyth2024naturallanguageguidancehighfidelity} only support basic tags (e.g. \textit{high-pitched, fast, female}) that can be extracted using signal processing tools; in contrast, we scale to a larger set of rich, abstract tags for the first time.

For intrinsic style tags, we use a perceptual speaker similarity model~\citep{ahn2024voxsimperceptualvoicesimilarity} to identify speakers whose speech resembles that of speakers human-annotated with intrinsic tags. Then, we propagate the intrinsic tags of the similar speaker, multiplying intrinsic data by $9$x to $2427$ hours. For situational style tags, we combine three different types of signals. We first identify expressive speech using an off-the-shelf dominance-valence-arousal speech classifier~\citep{wagner2023dawn}. Among the selected expressive speech clips, we use a text embedding model~\citep{SFRAIResearch2024} to find transcripts that semantically match the desired situational tag. Lastly, we use a large-scale speech-text multimodal LLM~\citep{geminiteam2024gemini15unlockingmultimodal} to check whether the speech acoustically matches the situational tag. We use these together to multiply situational data by $3$x to $215$ hours. 

We verify the quality of our collected data comprehensively. First, we perform human evaluation and show that annotators rate our automatically scaled data to be on par with human-annotated data in terms of adherence to the annotated style tags. Then, we train a style-prompted TTS model by finetuning the widely-used Parler-TTS~\citep{lacombe-etal-2024-parler-tts, lyth2024naturallanguageguidancehighfidelity} model on our dataset. We evaluate its performance in terms of speech style consistency, speech quality, and intelligibility. Our model shows significant gains in style consistency ($+7.9\%$ Consistency MOS) and quality ($+15.5\%$ Naturalness MOS) when compared to our best baseline finetuned on existing smaller-scaled datasets~\cite{koizumi2023librittsrrestoredmultispeakertexttospeech,nguyen2023expressobenchmarkanalysisdiscrete,richter2024earsanechoicfullbandspeech}. A system demo is available at \url{https://paraspeechcaps.github.io/}. In summary, our contributions are:
\begin{itemize}[noitemsep,leftmargin=10px]
    \item We introduce ParaSpeechCaps, a large-scale style-captioned dataset that covers $59$ unique style tags.
    \item We newly collect $282$ hours of crowdsourced intrinsic annotations for our human-annotated portion.
    \item We propose two novel approaches to automatically annotate rich style tags for the first time and scale to $2427$ hours of data.
    \item We show that human evaluators rate our scaled data to be on par with our human-labelled data, and that a style-prompted TTS model finetuned on it achieves the highest style consistency and naturalness.
    \item We provide detailed analyses on each of our dataset design choices to contextualize their contributions.
\end{itemize}

\section{Style Tag Taxonomy}
\label{subsec:tagtaxonomy}

\subsection{Our taxonomy and coverage}
We first provide an overview of the types of style tags we study. We define a style factor~\citep{jin2024speechcraft, guo2022promptttscontrollabletexttospeechtext,ando2024factorconditionedspeakingstylecaptioning} as a speech characteristic that one wants to control and a style tag as a word that selects a value for the style factor. For example, pitch, rhythm, emotion are style factors and \textit{\{deep, shrill\}, \{singsong, monotonous\} \{angry, scared\}} are style tags for each. We broadly classify style tags along two axes, intrinsic vs. situational and rich vs. basic.

\textit{Intrinsic} tags are tied to a speaker's identity and persist across their utterances (e.g. pitch, texture and accent), while \textit{situational} tags are utterance-level (e.g. emotion and expressivity). While intrinsic annotations can be obtained on a per-speaker basis, situational annotations must be obtained on a per-utterance basis. \textit{Basic} tags can be easily extracted using signal processing tools or simple classifiers, while \textit{rich} tags are subjective and often require human annotations.

\begin{figure}[h]
    \centering
    \includegraphics[width=\linewidth]{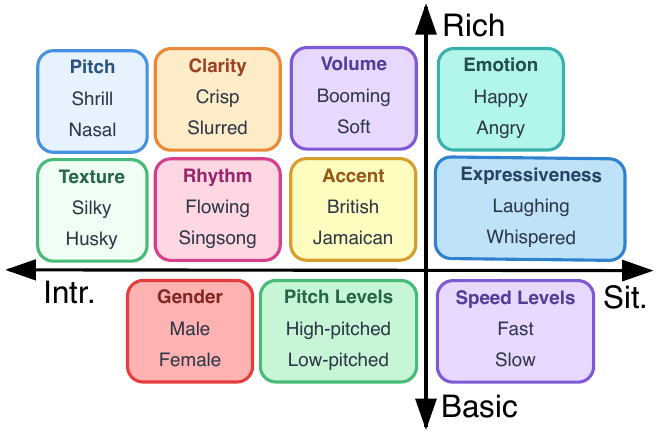}
    \caption{Our tag taxonomy that classifies along two axes, intrinsic (speaker-level) vs\@. situational (utterance-level) and rich (subjective) vs. basic (extractable via signal processing tools). Not all tags are shown; Appendix~\ref{app:taglist} has the full list of 59 tags.\vspace{-10pt}}
    \label{fig:tagtaxonomy}
\end{figure}

To comprehensively cover style types, we manually select $11$ style factors with an average of $5$ tags per style factor, resulting in $59$ total style tags consisting of $28$ rich intrinsic, $23$ rich situational and $5$ basic intrinsic and $3$ basic situational tags. Figure~\ref{fig:tagtaxonomy} visualizes our tag taxonomy with all $11$ style factors.

\subsection{Comparison to other datasets}
\label{subsec:datasetcomparisons}
Table~\ref{tab:datasetcomparisons} summarizes datasets from style-prompted TTS papers. We count the unique number of rich tags they support and dataset size (duration and speaker count). ParaSpeechCaps is the only large-scale, open-source dataset covering both rich intrinsic and situational tags.

\paragraph{Human-annotated datasets} InstructTTS (NLSpeech)~\citep{yang2023instructttsmodellingexpressivetts}, PromptStyle~\citep{liu2023promptstylecontrollablestyletransfer} and MEAD-TTS~\citep{guan2024mmttsmultimodalpromptbased} recruit humans to newly record or annotate emotional data, while TextrolSpeech~\citep{Ji2024} collates existing emotion datasets. These focus on $\approx 8$ emotions and some basic tags. Expresso~\citep{nguyen2023expressobenchmarkanalysisdiscrete} and EARS~\citep{richter2024earsanechoicfullbandspeech} cover a larger set of situational tags. LibriTTS-P~\citep{kawamura2024librittspcorpusspeakingstyle} collects intrinsic human annotations for LibriTTS-R~\citep{koizumi2023librittsrrestoredmultispeakertexttospeech}, while Coco-Nut~\citep{watanabe2023coco} collects diverse annotations.

\paragraph{Large-scale automatically scaled datasets} PromptTTS~\citep{guo2022promptttscontrollabletexttospeechtext} allows control over 4 emotions and is trained on a synthetic emotion dataset, PromptSpeech, generated via commercial TTS systems. While scalable, it only uses synthetic speech and is limited by the set of speakers and emotions supported by these TTS systems. PromptTTS2~\citep{leng2023prompttts2describinggenerating} largely focuses on an improved model architecture. Parler-TTS~\citep{lacombe-etal-2024-parler-tts, lyth2024naturallanguageguidancehighfidelity} proposes scaling up basic tags automatically using signal processing tools and rule-based binning. SpeechCraft~\citep{jin2024speechcraft} additionally uses an emotion classifier to scale $8$ emotions. AudioBox~\citep{vyas2023audioboxunifiedaudiogeneration} combines these approaches for scaling basic tags with human annotated rich tag datasets.
\begin{table}[h]
\centering
\small
\setlength{\tabcolsep}{3pt}
\begin{tabular}{lccccc}\toprule
&\multicolumn{3}{c}{\textbf{Rich}} & \multicolumn{2}{c}{\textbf{Size}} \\\cmidrule(lr){2-4}\cmidrule(lr){5-6}
\textbf{Dataset} &\textbf{I} &\textbf{S} & \textbf{\#} & \textbf{\#hr} & \textbf{\#spk} \\\midrule
\multicolumn{6}{l}{\small{\textit{\textbf{Open-Source}}}} \\
ParlerTTS\tiny{~\citep{lacombe-etal-2024-parler-tts}} &\xmark &\xmark & $0$ & $45$k & $8.0$k  \\
LibriTTS-R\tiny{~\citep{koizumi2023librittsrrestoredmultispeakertexttospeech}} &\xmark &\xmark & $0$ & $0.6$k & $2.4$k \\
PromptSpeech\tiny{~\citep{guo2022promptttscontrollabletexttospeechtext}} &\xmark &\cmark &4 & ? & $2.4$k \\
Expresso\tiny{~\citep{nguyen2023expressobenchmarkanalysisdiscrete}} &\xmark &\cmark & $18$ & $47$ & $4$ \\
EARS\tiny{~\citep{richter2024earsanechoicfullbandspeech}} &\xmark &\cmark & $18$ & $60$ & $107$ \\
TextrolSpeech\tiny{~\citep{Ji2024}} & \xmark &\cmark & $8$ & $0.3$k & $1.3$k  \\
MEAD-TTS\tiny{~\citep{guan2024mmttsmultimodalpromptbased}} &\xmark &\cmark & $8$ & $36$ & $47$  \\
SpeechCraft\tiny{~\citep{jin2024speechcraft}} &\xmark &\cmark & $7$ &$2.4$k &$5.9$k  \\
LibriTTS-P\tiny{~\citep{kawamura2024librittspcorpusspeakingstyle}} &\cmark &\xmark & $46$ &$0.6$k &$2.4$k  \\
Coco-Nut\tiny{~\citep{watanabe2023coco}} &\cmark &\cmark & ? & 8 & $7.3$k  \\
\textbf{ParaSpeechCaps (Ours)} &\cmark &\cmark & $51$ & $2.9$k & $45$k  \\\cmidrule(lr){1-6}
\multicolumn{6}{l}{\small{\textit{\textbf{Closed-Source}}}} \\
PromptTTS2\tiny{~\citep{leng2023prompttts2describinggenerating}} &\xmark &\xmark & 0 & $44$k & $7.5$k  \\
NLSpeech\tiny{~\citep{yang2023instructttsmodellingexpressivetts}} &\xmark &\cmark & ? & $44$ & $7$  \\
PromptStyle\tiny{~\citep{liu2023promptstylecontrollablestyletransfer}} &\xmark &\cmark & ? & $12$ & $6$  \\
AudioBox\tiny{~\citep{vyas2023audioboxunifiedaudiogeneration}} &\cmark &\cmark &? & ? &? \\
\bottomrule
\end{tabular}
\caption{A comparison of speech style-captioned datasets. Ours (ParaSpeechCaps) is the only large-scale open-source dataset that covers both rich intrinsic and situational tags. \textbf{Rich}: Rich tag support. \textbf{I}: Intrinsic, \textbf{S}: Situational, \textbf{\#}: Rich tag count. \textbf{\#hr}: Dataset duration. \textbf{\#spkr}: Speaker count. ?: unknown.\vspace{-10pt}}\label{tab:datasetcomparisons}
\end{table}

\begin{figure*}
    \centering
    \includegraphics[width=\linewidth]{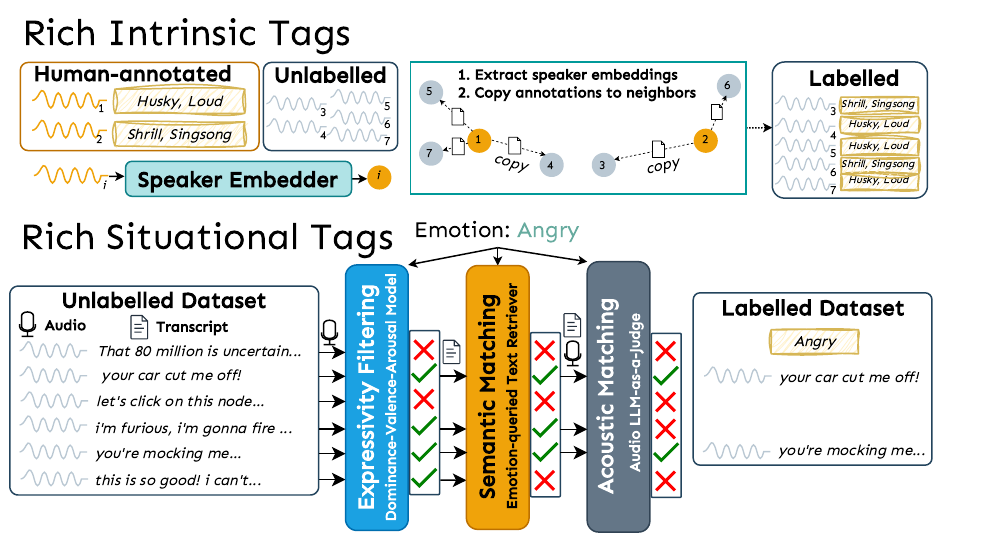}
    \caption{An overview of our automatic dataset scaling pipeline, for rich intrinsic and situational tags.\vspace{-10pt}}
    \label{fig:scaledfig}
\end{figure*}

\section{The ParaSpeechCaps Dataset}
\label{sec:datasets}

Our dataset aims to improve the \textbf{coverage of style tags} and provide ways to automatically gather \textbf{large-scale annotations} for rich tags without requiring human labor. We select a large set of $59$ style tags categorized by our taxonomy (Section~\ref{subsec:tagtaxonomy}), construct a human-annotated dataset (PSC-Base) covering all rich tags (Section~\ref{subsec:pscbase}) and develop our novel scalable annotation pipeline to create the PSC-Scaled dataset covering most rich tags (Section~\ref{subsec:pscscaled}), shown in Figure~\ref{fig:scaledfig}. All annotated style tags are converted to style prompts using a text LLM, Mistral-7B-Instruct-v0.2\citep{jiang2023mistral7b} (Appendix~\ref{app:promptgeneration}). 

\subsection{ParaSpeechCaps-Base}
\label{subsec:pscbase}

We hire Amazon Mechanical Turk workers to annotate speakers from Expresso~\citep{nguyen2023expressobenchmarkanalysisdiscrete}, EARS~\citep{richter2024earsanechoicfullbandspeech} consisting of enacted read speech and dialogue speech, as well as a $594$-speaker subset of VoxCeleb~\citep{nagrani2020voxceleb}) consisting of natural in-the-wild celebrity interviews.
The annotators provide all intrinsic tags in our ontology, excluding accent tags. We gather accent tags from metadata for Expresso and EARS and by prompting GPT-4 with the celebrity's name and ask it to output their accent for VoxCeleb (Appendix~\ref{app:promptgeneration}).

\paragraph{Annotator Qualification Task} We provide a simple task to annotators to check their ability to understand style tags, keeping only those $38$ that succeeded on at least 5 of 6 examples (Appendix~\ref{app:annotations}). 

\paragraph{Collecting Annotations} For each speaker, we create a single audio file consisting of multiple utterances ($3-8$ clips whose total duration is $20-40$ seconds). We provide this audio, the speaker's name (if available) and a list of our rich intrinsic tags with definitions and ask annotators to write at least $3$ distinct style tags. We collect $5$ annotations per speaker. Since this task is highly subjective, we keep only those tags that at least $2$ annotators agree on for our train and dev set, and only those that at least $3$ annotators agree on for our holdout set.

\paragraph{Selecting Speakers Representing Diverse Tags}
We identify celebrities, to annotate intrinsic speech tags for, by combining three sources: (a) an IMDb list~\citep{imdb_list}, (b) a ChatGPT-generated list of celebrities with distinctive voices and (c) the top $200$ longest Wikipedia pages for VoxCeleb celebrities (collected using~\citet{wikipedia_api}). This totals $302$ unique VoxCeleb celebrities. We collect annotations for them and find that the style tag distribution is imbalanced. For $12$ least-frequent tags~\footnote{lisp, hushed, pitchy, staccato, monotonous, punctuated, vocal fry, guttural, singsong, soft, stammering, shrill}, we use GPT-4~\citep{openai2024gpt4technicalreport} to obtain a list of celebrities that are likely to have them (details in Appendix~\ref{app:promptgeneration}), select a maximum of $40$ per tag, and end up with $187$ new celebrities to annotate. Finally, we randomly annotate $105$ additional celebrities, resulting in a total of $594$ celebrities.

\paragraph{{Supporting Rich Situational Tags}} We use Expresso~\citep{nguyen2023expressobenchmarkanalysisdiscrete} and EARS~\citep{richter2024earsanechoicfullbandspeech} annotated with speaking styles which we remap to our tag vocabulary. Table~\ref{tab:termremapping} in Appendix provides the full mapping of tags. For example, the \textit{fear} style is mapped to the tag \textit{scared}. Neutral speech and non-verbal sounds (e.g. coughing, yelling) are filtered out.

\paragraph{Generating Style Prompts} All annotated style tags are converted to style prompts using a text LLM, Mistral-7B-Instruct-v0.2\citep{jiang2023mistral7b} (Appendix~\ref{app:promptgeneration}). Since Expresso and EARS has both rich intrinsic and situational tag annotations, we generate two style prompts per example: one with only situational tags and one with both intrinsic and situational tags. Both style prompts are used when training. Since VoxCeleb has only intrinsic tag annotations, we generate one style prompt per example containing those tags. 

\paragraph{Train-Dev-Holdout Splits} We split PSC-Base into three splits called \textit{train}, \textit{dev} and \textit{holdout}; a tag-balanced subset of the \textit{holdout} split will eventually be our model evaluation dataset. For VoxCeleb, we find $64$ speakers that together ensure as far as possible that each rich intrinsic tag has $2$ male and $2$ female speakers available and place them into the holdout split. We place the remaining $530$ speakers into the train ($90\%$) and dev splits ($10\%$).  We place $80\%$ of Expresso in train, $10\%$ in dev and $10\%$ in holdout. We place unlabelled emotional utterances in EARS into the train set, and place the remaining utterances into train (80\%), dev (10\%), and holdout splits (10\%).  We ensure that there is no transcript overlap across splits, and in the case of VoxCeleb, no speaker overlap either.

\subsection{ParaSpeechCaps-Scaled}
\label{subsec:pscscaled}
We propose two approaches for scaling rich tag annotations, one for intrinsic tags and one for situational tags and apply both to the English portion of the large-scale Emilia~\citep{he2024emiliaextensivemultilingualdiverse} dataset (after preprocessing to remove infrequent speakers with $<5$ min) to create PSC-Scaled. All style factors except clarity and expressiveness are supported. We evaluate its quality and ablate design choices via human evaluation in Section~\ref{sec:datasetexps}.
\paragraph{{Scaling Intrinsic Tags}} Perceptual speaker similarity refers to how similar humans \textit{perceive} two speakers. This differs from standard speaker similarity rooted in speaker verification which measures the likelihood that two speakers are exactly the same. Based on initial manual analyses, we find that two speakers with high perceptual similarity usually share most intrinsic tags excluding clarity tags. For every human-annotated VoxCeleb speaker from PSC-Base and every Emilia speaker, we compute a median perceptual speaker embedding over 10 randomly-sampled utterances from that speaker using VoxSim~\citep{ahn2024voxsimperceptualvoicesimilarity}. For each VoxCeleb speaker, we find Emilia speakers that have a cosine similarity of at least $0.8$ (corresponding to a similarity rating of 5 out of 6 in VoxSim) and copy all intrinsic tags (excluding clarity tags) from the VoxCeleb speaker to these Emilia speakers.

\paragraph{{Scaling Situational Tags}} We encounter two major challenges in scaling situational tags: (a) \textbf{insufficient expressive data}: A major portion of an internet-scale speech dataset like Emilia is neutral and does not strongly exhibit emotions. (b) \textbf{no automatic classifiers:} There are no automatic classifiers covering all of our tags; classifiers such as emotion2vec~\citep{ma2023emotion2vecselfsupervisedpretrainingspeech} only support $8$ emotions. To solve the first challenge, we propose an \textbf{Expressivity Filtering} step to keep only highly expressive speech. To solve the second challenge, we propose a \textbf{Semantic Matching} step to find utterances that semantically match a desired emotion and an \textbf{Acoustic Matching} step to find utterances that acoustically match a desired emotion. Our overall pipeline cascades all three steps.

\begin{itemize}[noitemsep,leftmargin=10px]
\item \textbf{Expressivity Filtering}
The dominance-valence-arousal theory~\citep{RUSSELL1977273} posits that emotions live in a three-dimensional space consisting of dominance (degree of control), arousal (intensity) and valence (pleasantness), each with values between $0$ and $1$. Backed by~\citet{8003425}, we expect that utterances with extreme values for any one of these are likely to be expressive. Using an off-the-shelf DVA classifier~\citep{wagner2023dawn}, we filter for those utterances that have at least one value below $0.35$ or above $0.75$. We further filter using emotion-specific directions (e.g. for \textit{angry}, we expect the dominance or arousal to be high, and the valence to be low) (Appendix~\ref{sec:dvadirections}).

\item \textbf{Semantic Matching}
Recent work~\citep{chen2024emoknobenhancevoicecloning} shows that the speech transcript can be used to find utterances whose speaking style match a desired emotion. We embed speech transcripts from the Expressivity-Filtered dataset and queries of the form \textit{Instruct: Given an emotion, retrieve relevant transcript lines whose overall style/emotions matches the provided emotion. Query: \{emotion\}}
using a sentence embedding model (SFR-Embedding-Mistral~\citep{SFRAIResearch2024}) and sort by the cosine similarity between the query and the transcripts. Because the retriever overranks transcripts containing keywords related to the emotion (e.g. a transcript that contains the word \textit{angry} will be ranked even though it does not semantically convey the angry emotion), we filter transcripts that contain such emotion-specific keywords (Appendix~\ref{sec:dvadirections}).

\item \textbf{Acoustic Matching}
The semantic matching process results in many false positives. To filter these out, we take the top $100$k examples per emotion from the dataset sorted by the Semantic Matching step and prompt Gemini 1.5 Flash~\citep{geminiteam2024gemini15unlockingmultimodal}, a strong audio LLM, to rate on a $5$-point Likert scale whether the utterance matches the desired emotion, asking it to focus exclusively on the tone and not on the content (full prompt in Appendix~\ref{app:promptgeneration}). We keep only those examples that obtain a $5$ score.
\end{itemize}

\begin{figure*}
    \centering
    \begin{minipage}[t]{0.4909\linewidth}
        \includegraphics[width=\linewidth]{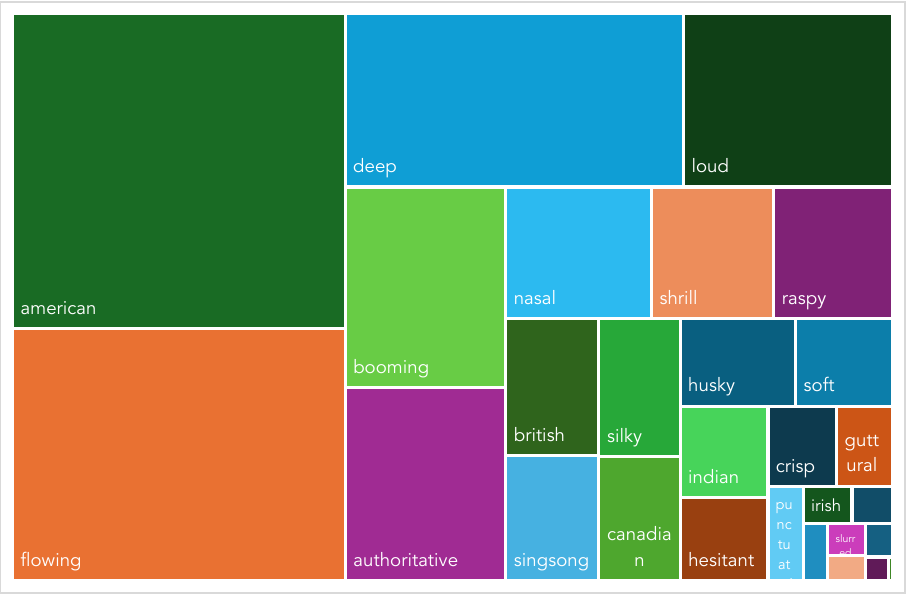}
        \label{fig:figure1}
    \end{minipage}
    \begin{minipage}[t]{0.4090\linewidth}
        \includegraphics[width=\linewidth]{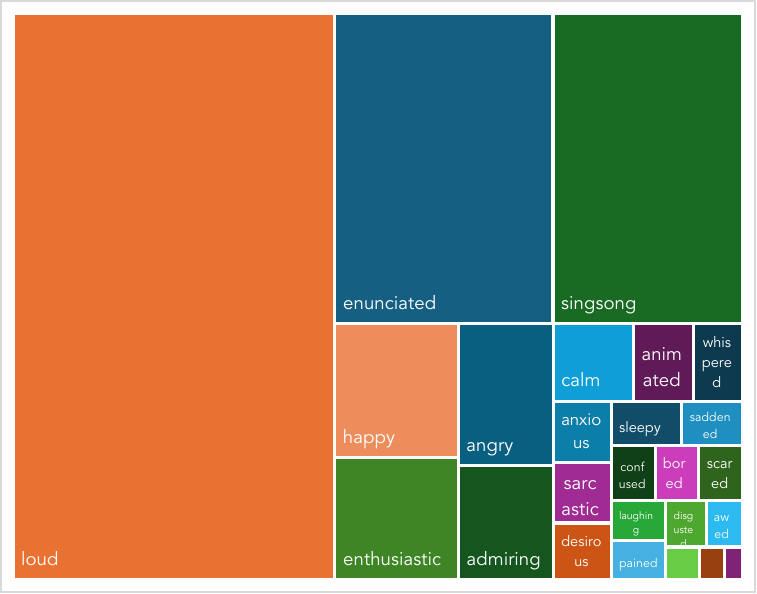}
    \end{minipage}
    \vspace{-10pt}
    \caption{Distribution of rich intrinsic (left, $2518$ hrs) and situational (right, $298$ hrs) tags in ParaSpeechCaps.\vspace{-10pt}}
    \label{fig:richtagstatistics}
\end{figure*}

\paragraph{Generating Style Prompts} All annotated style tags are converted to style prompts using a text LLM, Mistral-7B-Instruct-v0.2\citep{jiang2023mistral7b} (Appendix~\ref{app:promptgeneration}). We generate two style prompts per example that has both rich intrinsic and situational tag annotations: one with only intrinsic tags and one with both intrinsic and situational tags. Both style prompts are used when training. For all other examples that have either intrinsic or situational tags, we generate one style prompt per example.

\subsection{Extracting Basic Tags}
\label{subsec:basic}

We automatically annotate all data in ParaSpeechCaps with basic tags (gender, pitch levels and speed levels). Because much of our data has background noise, we also extract noise level tags ranging from \textit{very clear} to \textit{very noisy} to help the model separate noisy speech from clear speech; at inference, we use a \textit{clear} tag. 

\paragraph{{Gender}} We use dataset metadata for Expresso and EARS and prompt GPT-4 with the celebrity's name and ask it to output their gender for VoxCeleb (Appendix~\ref{app:promptgeneration}). For the rich intrinsic component of PSC-Scaled, we copy the gender tag of the parent VoxCeleb speaker to the Emilia speaker. For the rich situational component of PSC-Scaled, we apply a gender classifier~\citep{burkhardt2023speechbasedagegenderprediction} on a maximum of $50$ utterances per speaker and use the majority gender tag.

\paragraph{{Pitch, Speed and Noise Levels}}  For pitch, we use PENN~\citep{morrison2023cross} to compute the mean pitch across all utterances of a given speaker. We apply gender-dependent thresholds to label with \textit{low-, medium-} or \textit{high-pitched}. For speed, we use g2p~\citep{pine-etal-2022-gi22pi} to compute the number of phonemes per second and apply thresholds to label with \textit{slow, measured} or \textit{fast}. For noise levels, we use Brouhaha~\citep{lavechin2023brouhaha} to compute the signal-to-noise ratio and use Parler-TTS~\citep{lacombe-etal-2024-parler-tts}'s noise bins for the \textit{very noisy, quite noisy, slightly noisy, balanced in clarity, slightly clean, quite clean} and \textit{very clean} tags. All threshold values are available in Appendix~\ref{sec:thresholds}. We use the Dataspeech~\citep{lacombe-etal-2024-dataspeech} library.

\subsection{Dataset Statistics}
Figure~\ref{fig:richtagstatistics} showcases the distribution of different style tags in our ParaSpeechCaps dataset\footnote{We only provide textual annotations for existing datasets. Their speech data is subject to their own licenses.} (combining PSC-Human and PSC-Scaled).

\section{Verifying Scaled Data Quality}
\label{sec:datasetexps}
In this section, we provide human evaluation results for the scaled dataset we constructed in order to verify the quality of our automatic annotations. 

\subsection{Scaled Dataset Ablations}
We compare our initial human-annotated dataset (PSC-Base), our  automatically scaled dataset (PSC-Scaled) and ablated versions of PSC-Scaled, described below.

\paragraph{Rich Intrinsic Tags} We used a perceptual speaker embedding model, VoxSim~\citep{ahn2024voxsimperceptualvoicesimilarity}, to construct the intrinsic component of PSC-Scaled. We ablate it by creating a \textbf{Std. Embedder} version that uses a standard WavLM Large~\citep{chen2022wavlm} ECAPA-TDNN embedder. We select a cosine similarity threshold of $0.41$ that scales to approximately the same number of total speakers as PSC-Scaled.

\paragraph{Rich Situational Tags} 
We constructed the situational component of PSC-Scaled by pipelining three steps: \textbf{Expressivity Filtering}, \textbf{Semantic Matching} and \textbf{Acoustic Matching}. We create $3$ ablated versions that each skip one of these:
\begin{itemize}[noitemsep,leftmargin=10px]
    \item \textbf{w/o Expressivity Filtering} We apply Semantic and Acoustic Matching starting from the entire Emilia dataset without Expressivity Filtering.
    \item \textbf{w/o Semantic Matching} We run Acoustic Matching on random $100$k examples per emotion from the Expressivity-Filtered dataset.
    \item \textbf{w/o Acoustic Matching} We take the same number of examples per emotion as PSC-Scaled from the top of the Semantic Matching-sorted dataset without Acoustically Matching them.
\end{itemize}

\subsection{Evaluation Setup}
We recruit annotators on Amazon Mechanical Turk (Appendix~\ref{app:annotations}) collecting three annotations per example. We provide annotators a speech clip and its associated rich tag and ask them whether they hear it. For each tag, we compute its recall (fraction of instances in which it was selected) and report the average Tag Recall.

For each intrinsic tag, we sample a maximum of $12$ speakers and $4$ utterances per speaker for human evaluation (skipping $4$ tags: \textit{guttural, vocal-fry, monotonous, punctuated} as they have an insufficient number of speakers) from each dataset, totalling $356$, $420$ and $376$ examples for PSC-Base, PSC-Scaled and the Std. Embedder ablation respectively. For each situational tag, we randomly sample $20$ examples per emotion for human evaluation from each dataset, totalling $360$ examples per dataset.

\subsection{Main Results}

\begin{table}
\centering
\begin{tabular}{lcc}
\toprule
& \multicolumn{2}{c}{\textbf{Tag Recall} $\mathbf{\uparrow}$} \\\cmidrule(lr){2-3}
\textbf{Dataset} & \textbf{Intrinsic} & \textbf{Situational} \\
\midrule
PSC-Base & $48.7\%$ & $68.1\%$ \\
PSC-Scaled & $\mathbf{50.3\%}$ & $\mathbf{71.3\%}$ \\
\midrule
\multicolumn{3}{l}{\small{\textbf{\textit{Ablations}}}} \\
Std. Embedder & $45.3\%$ & -- \\
w/o Expressivity & -- & $61.0\%$ \\
w/o Semantic & -- & $66.1\%$ \\
w/o Acoustic & -- & $63.3\%$ \\
\bottomrule
\end{tabular}
\caption{Human evaluation of intrinsic/situational style tag recalls, comparing our datasets and ablations.\vspace{-10pt}}
\label{tab:datasetablations}
\end{table}

Table~\ref{tab:datasetablations} presents our evaluation results. For rich intrinsic tags, PSC-Scaled achieves a comparable performance to PSC-Base, while Std. Embedder worsens it. For rich situational tags, PSC-Scaled achieves a comparable performance to PSC-Base, while removing any of Expressivity Filtering, Semantic Matching, or Acoustic Matching worsens it. This shows that each step in our scaling pipeline is necessary and that it creates data of comparable quality to human annotations. In absolute terms, the tag recalls of PSC-Base are lower than $100\%$ which we attribute to human subjectivity for tag identification.

\section{Style-Prompted TTS Experiments}
In this section, we verify the utility of ParaSpeechCaps by using it to train style-Prompted TTS models.

\begin{table*}\centering
\begin{tabular}{lcccccc}\toprule
\setlength{\tabcolsep}{2pt}
&  \multicolumn{3}{c}{\textbf{Style Consistency}} & \multicolumn{1}{c}{\textbf{Quality}} & \multicolumn{2}{c}{\textbf{Intelligibility}} \\\cmidrule(lr){2-4}\cmidrule(lr){5-5}\cmidrule(lr){6-7}
\textbf{Model} & \textbf{CMOS} $\mathbf{\uparrow}$ & \textbf{Intr TR} $\mathbf{\uparrow}$ & \textbf{Sit TR} $\mathbf{\uparrow}$ & \textbf{NMOS} $\mathbf{\uparrow}$ & \textbf{IMOS} $\mathbf{\uparrow}$ & \textbf{WER} $\mathbf{\downarrow}$ \\\midrule
Ground Truth & $4.42 \scriptstyle{\pm 0.07}$ & $88.7\%$ & $88.6\%$ & $4.36 \scriptstyle{\pm 0.07}$ & $4.28 \scriptstyle{\pm 0.06}$ & $7.93$ \\
\cmidrule{1-7}
\small{\textit{\textbf{Baselines}}} & & & & & & \\
Parler-TTS & $3.05 \scriptstyle{\pm 0.08}$ & $33.0\%$ & $21.2\%$ & $2.85 \scriptstyle{\pm 0.07}$ & $4.31 \scriptstyle{\pm 0.07}$ & $4.62$ \\
+LTTSR & $3.07 \scriptstyle{\pm 0.08}$ & $33.7\%$ & $22.4\%$ & $2.95 \scriptstyle{\pm 0.07}$ & $\mathbf{4.44} \scriptstyle{\pm 0.06}$ & $\mathbf{4.47}$ \\
+LTTSP,Exp,EARS & $3.55 \scriptstyle{\pm 0.08}$ & $40.7\%$ & $69.7\%$ & $3.10 \scriptstyle{\pm 0.07}$ & $4.19 \scriptstyle{\pm 0.07}$ & $7.14$ \\
\cmidrule{1-7}
\small{\textit{\textbf{Our Models}}} & & & & & & \\
\textbf{Base}: +VoxC,Exp,EARS & $3.75 \scriptstyle{\pm 0.08}$ & $63.6\%$ & $68.1\%$ & $3.27 \scriptstyle{\pm 0.08}$ & $4.05 \scriptstyle{\pm 0.07}$ & $9.14$ \\
\textbf{Scaled:} +VoxC,Exp,EARS,Emilia & $\mathbf{3.83} \scriptstyle{\pm 0.08}$ & $\mathbf{69.5\%}$ & $\mathbf{75.4\%}$ & $\mathbf{3.58} \scriptstyle{\pm 0.07}$ & $4.07 \scriptstyle{\pm 0.07}$ & $8.63$ \\
\bottomrule
\end{tabular}
\caption{Evaluation results comparing style consistency (CMOS, Intrinsic and Situational Rich Tag Recall), speech quality (NMOS) and intelligibility (IMOS, WER). Mean score and 95\% confidence intervals are reported for MOS. Our Base and Scaled models obtain improved style consistency (+5.6\% and +7.9\% Consistency MOS) and speech quality (+5.5\% and +15.5\% Naturalness MOS) over baselines.\vspace{-10pt}}\label{tab:results}
\end{table*}

\subsection{Experimental Setup}

\paragraph{{Main Evaluation Dataset}} We create a tag-balanced test dataset consisting of $246$ examples from the \textit{holdout} split of PSC-Base (Section~\ref{subsec:pscbase}) that evaluates adherence to one rich tag at a time. For each tag, we select a maximum of five clips, covering as many speakers as possible. Then, for each clip, we construct a tag set consisting of the rich tag, one to three basic tags (we always include gender, and randomly include pitch and speed with a $50\%$ probability), and a \textit{clear} noise tag, and convert to style prompts.

\paragraph{{Compositional Evaluation Dataset}} We create a compositional style prompt dataset that evaluates simultaneous adherence to two rich tags (one intrinsic, one situational). We select $12$ intrinsic tags (\textit{shrill, deep, husky, guttural, soft, authoritative, crisp, slurred, hesitant, flowing, british, canadian}), randomly select $10$ situational tags (\textit{desirous, animated, sarcastic, pained, admiring, whispered, awed, anxious, enunciated, sleepy}) and use both genders (\textit{male, female}) creating $12 \times 10 \times 2 = 240$ compositions. We sample $240$ random transcripts of $6-10$ words from the LibriTTS test set. Note that is no ground truth speech for these compositional examples.

\paragraph{{Evaluation Metrics}}
We evaluate for style consistency (Consistency MOS, Tag Recall), speech quality (Naturalness MOS),  and intelligibility (Intelligibility MOS, WER). Except WER, all other metrics rely on human evaluation due to lack of robust automatic evaluation metrics, in line with prior work. For human evaluation, we recruit annotators on Amazon Mechanical Turk (details in Appendix~\ref{app:annotations}), collect 3 annotations per example and report the mean and $95\%$ confidence intervals for MOS~\citep{5946971}.

\begin{itemize}[noitemsep,leftmargin=10px]
    \item \textbf{Style Consistency} We report CMOS (Consistency MOS) where each annotator is asked to rate the agreement between a given speech clip and the style prompt on a 5-point Likert scale, similar to~\citet{vyas2023audioboxunifiedaudiogeneration}. Since the style prompt contains a mix of rich and basic tags, for our main evaluation, we additionally ask annotators to select whether they specifically hear the \textbf{rich} tag for a more finegrained evaluation. For each rich tag, we compute its recall (fraction of instances in which it was selected), and report the average Tag Recall over intrinsic and situational tags separately. For the compositional evaluation experiment that contain both intrinsic and situational tags, we instead assess whether the model generated both types of tags, just intrinsic, just situational or neither.
    \item \textbf{Quality} We report NMOS (Naturalness MOS) where each annotator is asked to rate the naturalness and realisticity of a given speech clip on a 5-point Likert scale, similar to~\citet{vyas2023audioboxunifiedaudiogeneration}.
    \item \textbf{Intelligibility} We report IMOS (Intelligibility MOS) where each annotator is asked to rate the intelligibility of a given speech clip on a 5-point Likert scale, similar to~\citet{peng2024voicecraftzeroshotspeechediting}. We report a text-normalized Word Error Rate (WER) between the ASR transcript of the clip and the input transcript using distil-whisper/distil-large-v2~\citep{gandhi2023distilwhisperrobustknowledgedistillation} and the Whisper text normalizer.
\end{itemize}

\paragraph{{Model Architecture}} We use Parler-TTS~\citep{lyth2024naturallanguageguidancehighfidelity,lacombe-etal-2024-parler-tts},\footnote{parler-tts/parler-tts-mini-v1 checkpoint.} an $880$M parameter style-prompted TTS model trained on Librispeech~\citep{Pratap_2020} and LibriTTS-R~\citep{koizumi2023librittsrrestoredmultispeakertexttospeech} that can control pitch, speed, gender and expressivity style factors. We briefly describe its architecture here; it has two main components: the Parler-TTS decoder LM that autoregressively generates DAC~\citep{kumar2023highfidelityaudiocompressionimproved} audio tokens, and a frozen text encoder, Flan-T5-Large~\citep{chung2022scaling}. The style prompt is encoded by this text encoder and made available to the decoder LM via cross-attention. The text transcript is tokenized by Flan-T5 and prefilled to the decoder LM.

\paragraph{Dataset Sampling} PSC-Scaled is $\approx 8$x larger than PSC-Base, and intrinsically tagged data is $\approx 8$x larger than situationally tagged data. To reduce dataset imbalance during training, for all our models and baselines, we upsample VoxCeleb data by $2$x, Expresso and EARS data by $6$x, and the situational component of PSC-Scaled by $2$x when training.

\paragraph{{Inference Setup}} We perform inference using temperature $1.0$, repetition penalty $1.0$ and a maximum of $2580$ tokens. Because autoregressive TTS inference is unstable~\citep{han2024vallerrobustefficient}, we sample a maximum of $3$ times, stopping when the sample's WER $<20$ and selecting the sample with the lowest WER otherwise. Although we do not train with classifier free guidance~\citep{ho2022classifierfreediffusionguidance} we find that including it at inference with a $1.5$ scale consistently improves style consistency (Section~\ref{subsec:discussion}) and do so for all models. We represent the unconditional prompt as a zero-tensor.

\subsection{Comparison Systems}

\paragraph{Our models}
We train a \textit{Base} model on the train set of PSC-Base (VoxCeleb, Expresso and EARS) and a \textit{Scaled} model combining PSC-Base and PSC-Scaled. Since Parler-TTS is trained on LibriTTS-R, we include a $150$-hr random subset of LibriTTS-R train set annotated with basic tags for regularization. We train both models with a total batch size of $32$, a weight decay of $0.01$ and cosine schedulers with no warmup. We train our Base model on 4 NVIDIA A40 GPUs for $140$k steps with a peak LR of $8 \times 10^{-5}$, and use the same configuration for all baselines. We train our Scaled model on 4 NVIDIA H100 GPUs for $840$k steps in $2$ $420$k-step stages: a first stage with a peak LR of $8 \times 10^{-5}$ and a second stage with a peak LR of $4 \times 10^{-5}$ initialized from the first stage. As PSC-Scaled is much larger than PSC-Base, we train the model for longer.

\paragraph{Parler-TTS} We initialize all baselines and our models with the Parler-TTS-Mini-v1 model, denoted Parler-TTS.
\paragraph{+LTTSR} We finetune Parler-TTS on the LibriTTS-R~\citep{koizumi2023librittsrrestoredmultispeakertexttospeech} dataset annotated with basic tags. This baseline ablates training on only basic tags vs. rich tags for the same number of steps.
\paragraph{+LTTSP,Exp,EARS} We train with LibriTTS-P~\citep{kawamura2024librittspcorpusspeakingstyle}, a dataset that annotates LibriTTS-R with a different set of rich intrinsic tags, combined with Expresso and EARS with rich situational tags. LibriTTS-P provides three annotations per speaker and each style tag may have strength qualifiers (\textit{slightly, very}). We remove \textit{slightly} tags and remap some to our vocabulary (see Appendix~\ref{app:preprocessing}). We randomly select one of the three annotations and extract basic tags ourselves. This baseline ablates our PSC-Base intrinsic tag data against LibriTTS-P.

\subsection{Main Results}
\label{subsec:results}
Table~\ref{tab:results} presents our results, comparing models for style consistency, speech quality and intelligibility. Our Scaled model achieves the highest style consistency, with clear improvements for both intrinsic and situational tags, as well as the highest naturalness.

\paragraph{{Speech-Style Consistency}} The low Consistency MOS and Tag Recalls of the Parler-TTS and +LTTSR models show that training on basic tags does not generalize to rich styles. Our Base model and the +LTTSP,Exp,EARS model is trained on the same situational tag data but different intrinsic tag data. Therefore, both models achieve similar Situational Tag Recalls but our model vastly improves Intrinsic Tag Recall ($40.7\% \rightarrow 63.6\%$), demonstrating that our human-annotated intrinsic data is superior in quality. Our Scaled model achieves even higher Consistency MOS ($3.73 \rightarrow 3.83$) and Tag Recalls (Intr: $63.6\% \rightarrow 69.5\%$, Sit: $68.1\% \rightarrow 75.4\%$) compared to our Base model, showing the benefit of scaling the dataset.

\paragraph{{Speech Quality}}
+LTTSP,Exp,EARS improves naturalness as compared to Parler-TTS and +LTTSR ($2.95 \rightarrow 3.10$), showing the benefits of training on existing rich style datasets. Our model trained on our human-annotated data (PSC-Base) further improves it ($  3.10 \rightarrow 3.27$) and training on PSC-Scaled vastly improves it ($3.27 \rightarrow 3.58$), again showcasing its utility.

\paragraph{{Intelligibility}}
Baselines trained on clean audiobook data with basic tags (Parler-TTS and +LTTSR) obtain the highest intelligibility MOS and lowest WER, both outperforming even the ground truth. Because these baselines generate neutral, non-expressive speech, they are easier to understand by both humans (IMOS) and ASR models (WER) as compared to the ground truth, while our models as well as the +LTTSP,Exp,EARS baseline trained on rich style data obtain a lower MOS score. We dig deeper into this result in Section~\ref{subsec:discussion}. We find that the largest gaps in intelligibility are caused by our models faithfully adhering to certain style tags that \textit{should} be naturally less intelligible to evaluators (e.g. non-American accents, clarity tags like \textit{slurred, stammering}, etc.).

\subsection{Compositionality Results}
\begin{figure}
\includegraphics[width=\linewidth]{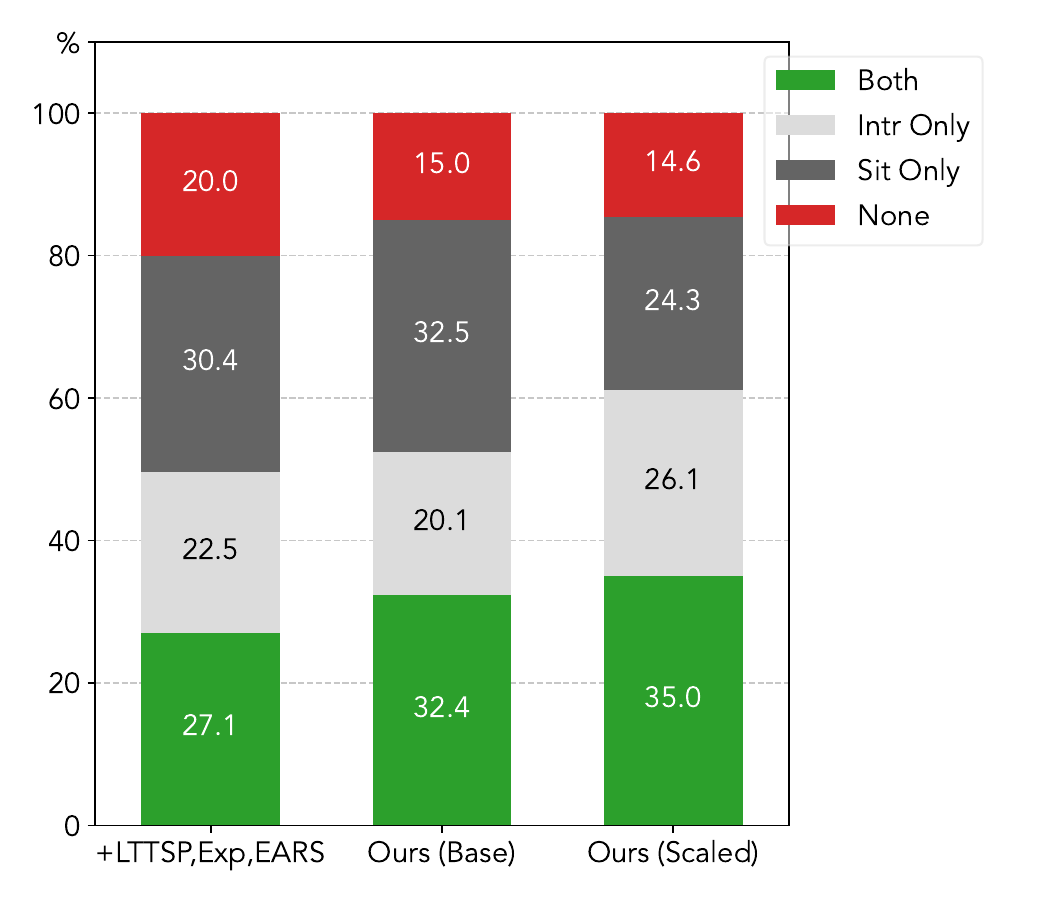}
\caption{Evaluation results for compositional style prompts. We report how frequently both types of tags, one of the two, or neither are generated. Our Scaled model achieves the highest compositionality.\vspace{-10pt}}
\label{fig:compositionalresults}
\end{figure}

Figure~\ref{fig:compositionalresults} presents our compositional evaluation results, where we present style prompts that simultaneously contain an intrinsic tag and a situational tag. We compare the best baseline (+LTTSP,Exp,EARS) with our Base and Scaled models. We find that our Scaled model correctly generates both tags more frequently than our Base model, which in turn outperforms the +LTTSP,Exp,EARS baseline. We also observe that when the models partially succeed by generating one of the two types, +LTTSP,Exp,EARS and our Base model prefer generating the situational tag, while our Scaled model prefers the intrinsic tag, likely owing to the large intrinsic component of PSC-Scaled.

\subsection{Discussion}
\label{subsec:discussion}
\paragraph{Why do models trained on rich style data have lower intelligibility?} We compute the difference in the Intelligibility MOS obtained by our Scaled model and the +LTTSR baseline, as well as the difference in the Tag Recall, broken down by tag. We find that amongst the top tags with the largest drop in IMOS, we find non-American accent tags (\textit{Indian, Scottish, Jamaican, Canadian}), clarity tags (\textit{slurred, stammering}), extreme emotions (\textit{pained}) which are naturally less intelligible to MTurk annotators (full results in Figure~\ref{fig:intelligibilityvsrecall} in the appendix). As shown by the Tag Recall difference, our model generates these tags more faithfully, and thus incurs this natural intelligibility drop, as compared to the +LTTSR baseline.

\paragraph{Inference-time classifier-free guidance improves style consistency, even without dropout-based training} Table~\ref{tab:cfgresults} presents human evaluation results for style consistency (Consistency MOS, Intrinsic and Situational Tag Recalls) using our main evaluation dataset, comparing models inferred with and without classifier-free guidance. Even though we do not train the model to handle empty style prompts using CFG dropout~\citep{ho2022classifierfreediffusionguidance} as is commonly done, we still find that all models are able to utilize it to improve style consistency across all metrics.

\begin{table}\centering
\small
\begin{tabular}{@{}l@{\hspace{4pt}}c@{\hspace{4pt}}c@{\hspace{4pt}}c@{\hspace{4pt}}c@{}}\toprule
\textbf{Model} & \textbf{CFG?} & \textbf{CMOS} $\mathbf{\uparrow}$ & \textbf{Intr TR} $\mathbf{\uparrow}$ & \textbf{Sit TR} $\mathbf{\uparrow}$ \\\midrule
\multirow{2}{*}{+LTTSP,Exp,EARS} &\xmark & $3.50 \scriptstyle{\pm 0.09}$ & $49.8\%$ & $66.7\%$\\
&\cmark & $\mathbf{3.64} \scriptstyle{\pm 0.10}$ & $\mathbf{51.2}\%$ & $\mathbf{73.3}\%$\\
\cmidrule{1-5}
\multirow{2}{*}{\textbf{Base} (Ours)} & \xmark & $3.76 \scriptstyle{\pm 0.09}$ & $67.1\%$ & $68.6\%$\\
& \cmark & $\mathbf{3.81} \scriptstyle{\pm 0.09}$ & $\mathbf{68.8}\%$ & $\mathbf{71.3}\%$\\
\cmidrule{1-5}
\multirow{2}{*}{\textbf{Scaled} (Ours)} & \xmark & $3.69 \scriptstyle{\pm 0.09}$ & $64.8\%$ & $65.1\%$\\
& \cmark & $\mathbf{3.92} \scriptstyle{\pm 0.08}$ & $\mathbf{70.7}\%$ & $\mathbf{76.4}\%$\\
\bottomrule
\end{tabular}
\caption{Human evaluation results ablating inference-time classifier-free guidance (CFG). We compare Consistency MOS and Intrinsic/Situational tag recall with and without inference-time classifier-free guidance (CFG). Mean score and 95\% confidence intervals shown for MOS. CFG improves style consistency across all metrics and models.}\label{tab:cfgresults}
\end{table}

\section{Related Work}
\paragraph{Style-Prompted Text-to-Speech Models} We describe style-prompted TTS papers in detail in Section~\ref{subsec:datasetcomparisons}. An orthogonal line of work~\citep{chen2024stylefusionttsmultimodalstylecontrol, zhu2024unistyle,yamamoto2024descriptionbasedcontrollabletexttospeechcrosslingual} innovates on style control architecture.

\paragraph{Style Control for other Speech Tasks}
Recent work has explored style prompts for tasks other than TTS. DreamVoice~\citep{hai2024dreamvoicetextguidedvoiceconversion} annotates LibriTTS-R with rich intrinsic tags for voice conversion. VCTK-RVA~\citep{sheng2024voiceattributeeditingtext} annotates the VCTK dataset with intrinsic tags for training a style-prompted speech editing system.

\section{Conclusion}
We present ParaSpeechCaps, a large-scale speech style captioned dataset that supports a rich and diverse set of styles covering both speaker-level intrinsic and utterance-level situational tags. Using our novel two-pronged scaling approach for intrinsic and situational tags, we create $2427$ hours of automatically annotated data, in addition to $282$ hours of human-labelled data. Our automatically annotated data quality is verified by human evaluators to be on par with human-labelled data. Furthermore, style-prompted TTS models finetuned on ParaSpeechCaps achieve the highest style consistency and naturalness as compared to baselines, showing its utility.

\section*{Acknowledgements}
We thank Puyuan Peng, Atula Tejaswi and other members of the UT NLP community for useful feedback. This work was done in part while the last author was visiting the Simons Institute for the Theory of Computing. We gratefully acknowledge use of the research computing resources of the Empire AI Consortium, Inc, with support from the State of New York, the Simons Foundation, and the Secunda Family Foundation.

\section*{Limitations}
\paragraph{Language coverage} We limit our current experiments to English data; there is a lot of potential to expand style-prompted TTS to more languages, both in terms of the language of the utterance and the language of the style prompt. Some work~\citep{jin2024speechcraft,yamamoto2024descriptionbasedcontrollabletexttospeechcrosslingual} explores  other languages like Chinese and Japanese in addition to English for style-prompted TTS.

\paragraph{Dataset biases} Our dataset creation methodology inevitably reflects and amplifies implicit correlations between tags, leading to potential coverage gaps. While some of these correlations are acoustically intuitive and expected (e.g., between \textit{high-pitched} and \textit{shrill} voices), others may perpetuate undesirable biases, particularly when style tags correlate with demographics. For example, we observed that the \textit{shrill} tag is overrepresented by female speakers, while the \textit{guttural} tag is overrepresented among male speakers. A trained model may learn these associations, potentially limiting its ability to generate diverse combinations of styles and speaker identities. Mitigating these biases is an important avenue for future work.

\paragraph{Lack of automatic metrics} This field requires expensive and subjective human evaluation metrics due to the lack of automatic evaluation, which prevents quick experimental turnarounds, large-scale evaluation datasets, and the ability to analyze model behavior in a finegrained manner. Future work can investigate how to develop automatic metrics for style-prompted TTS.

\bibliography{anthology,custom}

\appendix
\section{List of Speech Style Tags}
\label{app:taglist}
This is the list of tags we consider:
\begin{itemize}
    \item \textbf{Intrinsic:}
    \begin{itemize}
        \item \textbf{Rich:}
        \begin{itemize}
            \item \textbf{Pitch}: Shrill, Nasal, Deep.
            \item \textbf{Texture:} Silky, Husky, Raspy, Guttural, Vocal-fry.
            \item \textbf{Clarity:} Crisp, Slurred, Stammering.
            \item \textbf{Volume:} Booming, Authoritative, Loud, Soft.
            \item \textbf{Rhythm:} Flowing, Monotonous, Punctuated, Hesitant, Singsong.
            \item \textbf{Accent:} American, British, Scottish, Canadian, Australian, Irish, Indian, Jamaican.
        \end{itemize}
        \item \textbf{Basic:}
        \begin{itemize}
            \item \textbf{Pitch Levels:} High-pitched, Medium-pitched, Low-pitched.
            \item \textbf{Gender:} Male, Female.
        \end{itemize}
    \end{itemize}
    \item \textbf{Situational:}
    \begin{itemize}
        \item \textbf{Rich:}
        \begin{itemize}
            \item \textbf{Emotion:} Enthusiastic, Happy, Angry, Saddened, Awed, Calm, Anxious, Disgusted, Scared, Confused, Bored, Sleepy, Pained, Guilt, Sarcastic, Sympathetic, Admiring, Desirous.
            \item \textbf{Expressiveness:} Animated, Laughing, Passive, Whispered, Enunciated.
        \end{itemize}
        \item \textbf{Basic:}
        \begin{itemize}
            \item \textbf{Speed Levels:} Fast, Measured, Slow.
        \end{itemize}
    \end{itemize}
\end{itemize}

Some style factors like volume, speed and rhythm can technically be both intrinsic and situational. However, since we collect data for volume and rhythm with intrinsic human annotations, but extract speed tags on an utterance-level i.e. situationally, we place them in their respective categories. Manually written definitions for each style tag can be found in Table~\ref{tab:tagdefinitions}.
\begin{table*}[!htp]\centering
\scriptsize
\begin{tabular}{p{0.2\linewidth}p{0.7\linewidth}}\toprule
Attribute &Description \\\midrule
High-pitched &A voice with a distinctly high frequency. \\
Shrill &A high-pitched, piercing, and sharp voice. \\
Nasal &A whiny voice that sounds like someone is speaking through their nose. \\
Medium-pitched &A voice with a medium frequency that is neither very high or low-pitched. \\
Low-pitched &A voice with a distinctly low frequency. \\
Deep &A low-pitched, resonant, rich voice. \\
Silky &A smooth, pleasant and soothingly soft voice. \\
Husky &A slightly rough, low voice that conveys a gritty texture. \\
Raspy &A rough, grating, somewhat harsh voice. \\
Guttural &A deep, throaty, gravelly voice. \\
Vocal-fry &A creaky, breathy voice that occurs when vocal cords flutter and produce a sizzling, popping sound at ends of sentences. \\
American &A voice with an American accent. \\
British &A voice with a British accent. \\
Scottish &A voice with a Scottish accent. \\
Canadian &A voice with a Canadian accent. \\
Australian &A voice with a Australian accent. \\
Irish &A voice with an Irish accent. \\
Indian &A voice with an Indian accent. \\
Jamaican &A voice with an Jamaican accent. \\
Male &A male voice, often having a lower pitch. \\
Female &A female voice, often having a higher pitch. \\
Booming &A loud, resonant, commanding, powerful voice. \\
Authoritative &A confident, clear voice with a tone that conveys expertise and assurance. \\
Loud &A voice with a high volume. \\
Soft &A gentle, low-volume, calm and soothing voice typically used to convey subtlety. \\
Whispered &A breathy, low-volume voice typically used to speak discreetly. \\
Crisp &A clear, distinct, articulate voice. \\
Slurred &An unclear, difficult-to-understand voice that blends together sounds and words. \\
Stammering &A voice with pauses, repetitions and prolongations of words that disrupt the speech flow. \\
Singsong &A melodious voice that rises and falls in a musical manner. \\
Flowing &A clear, coherent, seamless and easy-to-understand voice. \\
Monotonous &A dull, flat voice whose pitch, tone and speed remains constant throughout. \\
Punctuated &An engaging voice with clear, deliberate pauses that emphasize key words. \\
Enunciated &A voice that clearly and precisely articulates words, with each syllable distinctly pronounced. \\
Fast speed &A rapidly speaking, quick voice with few pauses. \\
Measured speed &A controlled, deliberate voice that has an even tone and a moderate speed. \\
Slow speed &A voice with a slower speaking rate. \\
Hesitant &An uncertain, tentative voice, often marking a lack of confidence, reluctance or confusion. \\
Enthusiastic &A lively, energetic, positive voice that conveys excitement and interest in the topic being discussed. \\
Happy &A warm, positive and joyful voice. \\
Angry &A raised voice that conveys anger, frustration or displeasure, characterized by raised volume and emphatic speech patterns. \\
Saddened &A voice with a low, subdued, and unenergetic tone that conveys distress, disappointment or sadness. \\
Awed &A voice that conveys the speaker's admiration, wonder or reverance of something the speaker appreciates. \\
Calm &A calm, gentle and serene voice that conveys the speaker's relaxed and peaceful emotion. \\
Anxious &A voice that conveys nervousness and anxiety, often marked by rapid or jittery speech patterns. \\
Disgusted &A intonated voice that conveys repulsion and disgust by appropriately altering its pitch and rhythm. \\
Scared &A shaky, rapid voice that reflects the speaker's anxiety or fear. \\
Confused &A voice characterized by indecision and a lack of clarity, often marked by hesitance. \\
Bored &A voice, often monotonous, that indicates lack of enthusiasm and disinterest. \\
Sleepy &A soft, slow, low-energy voice that indicates tiredness. \\
Pained &A voice characterized by a strained, trembling tone that indicates sorrow or anguish. \\
Guilt &A voice that carries a wavering, hesitant tone that hints at discomfort or regret. \\
Sarcastic &A speaking style that is characterized by a distinct tone of irony that suggests that the speaker's intention is to mock or convey contempt. \\
Sympathetic &A gentle, compassionate voice that reassures and seeks to empathize with the listener. \\
Admiring &An appreciative, positive and complimentary manner of speaking. \\
Desirous &An emotional voice that conveys deep longing or desire. \\
Animated &A energetic, heightened voice characterized by varied intonations or emotional intensity. \\
Laughing &A voice with intermittent sounds of laughter conveying amusement and joy. \\
Passive &A tentative, subdued and uninterested voice. \\
\bottomrule
\end{tabular}
\caption{Manually written style tag definitions.}\label{tab:tagdefinitions}
\end{table*}

\begin{figure*}
    \centering
    \includegraphics[width=0.8\linewidth]{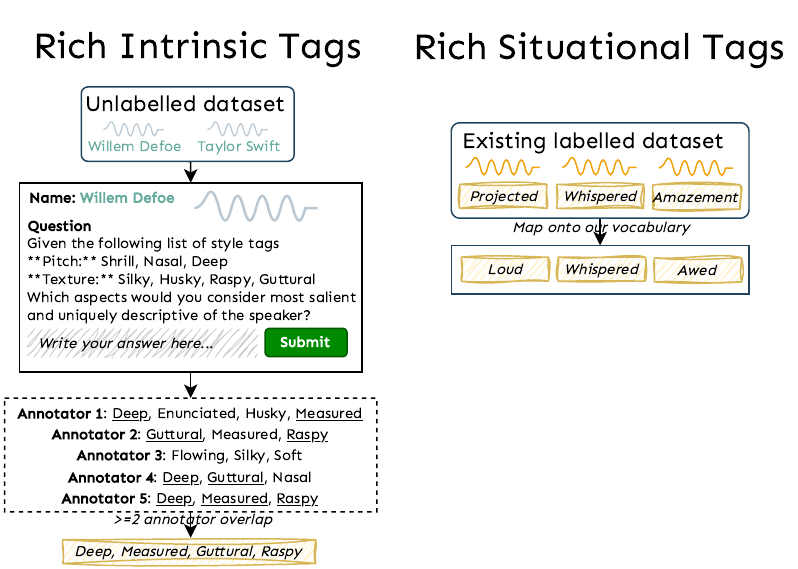}
    \caption{An overview of our human annotation pipeline, for rich intrinsic and situational tags.}
    \label{fig:humanfig}
\end{figure*}

\section{Human Annotation: Details}
\label{app:annotations}
We visualize our human annotation pipeline in Figure~\ref{fig:humanfig}.
\subsection{Annotation Details}
We recruit Amazon Mechanical Turk workers with a Masters certification with a minimum approval rate of $99\%$ and at least $5000$ successful HITs situated in the United States. For training dataset annotations, we perform a qualification task using $6$ pairs of manually selected clips from VoxCeleb or Expresso where one clip exhibits a style (one of \textit{deep, whispered, scared, slurred, high-pitched, enunciated}) and the other doesn't, and select $38$ annotators that succeeded on at least $5$. We pay $\$9$/hr.

\subsection{Annotation User Interfaces}
\begin{figure*}[htp!]
    \centering
    \includegraphics[width=\linewidth]{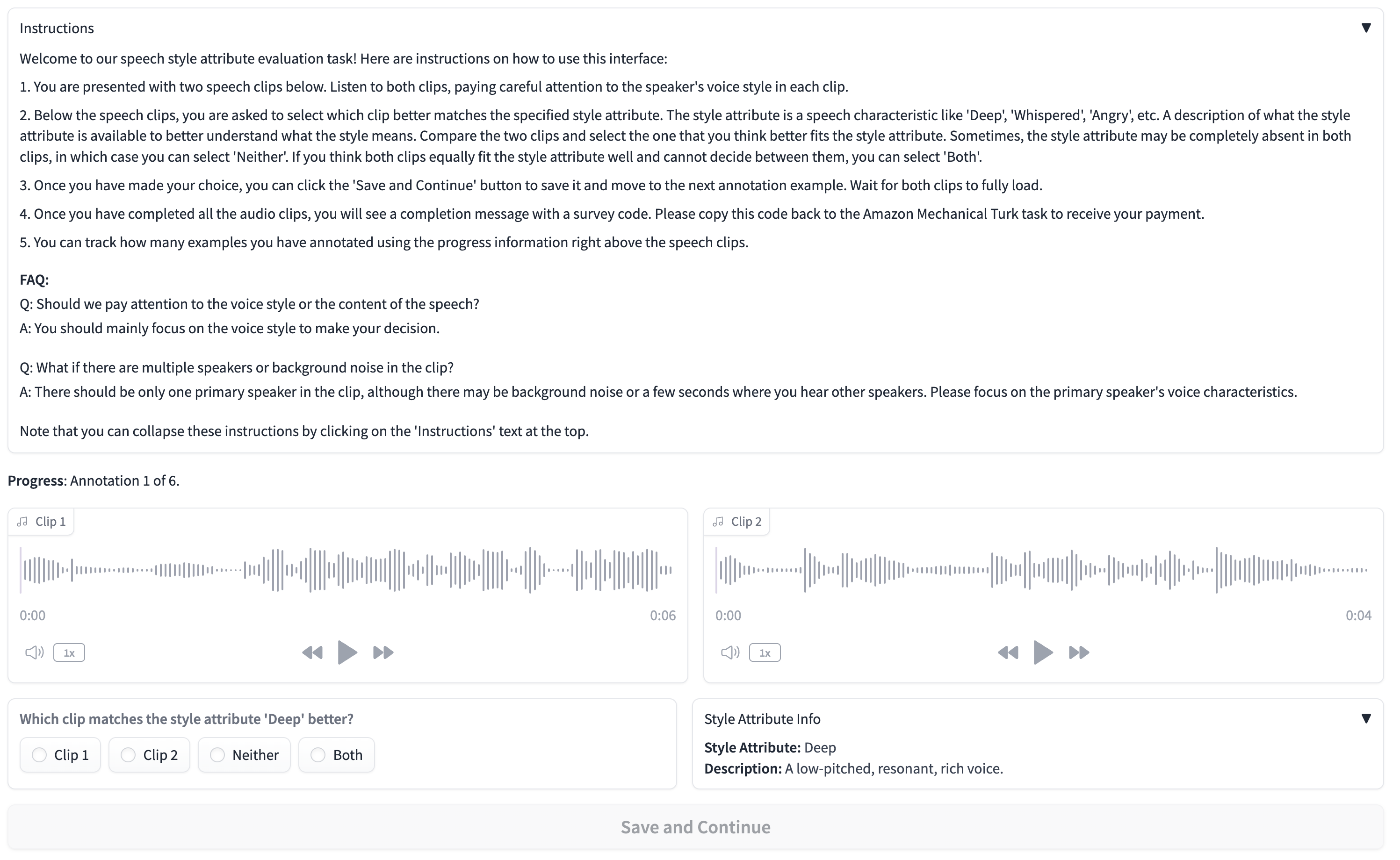}
    \caption{Annotation UI for selecting qualified annotators.}
    \label{fig:qualification_task_UI}
\end{figure*}

\begin{figure*}[htp!]
    \centering
    \includegraphics[width=\linewidth]{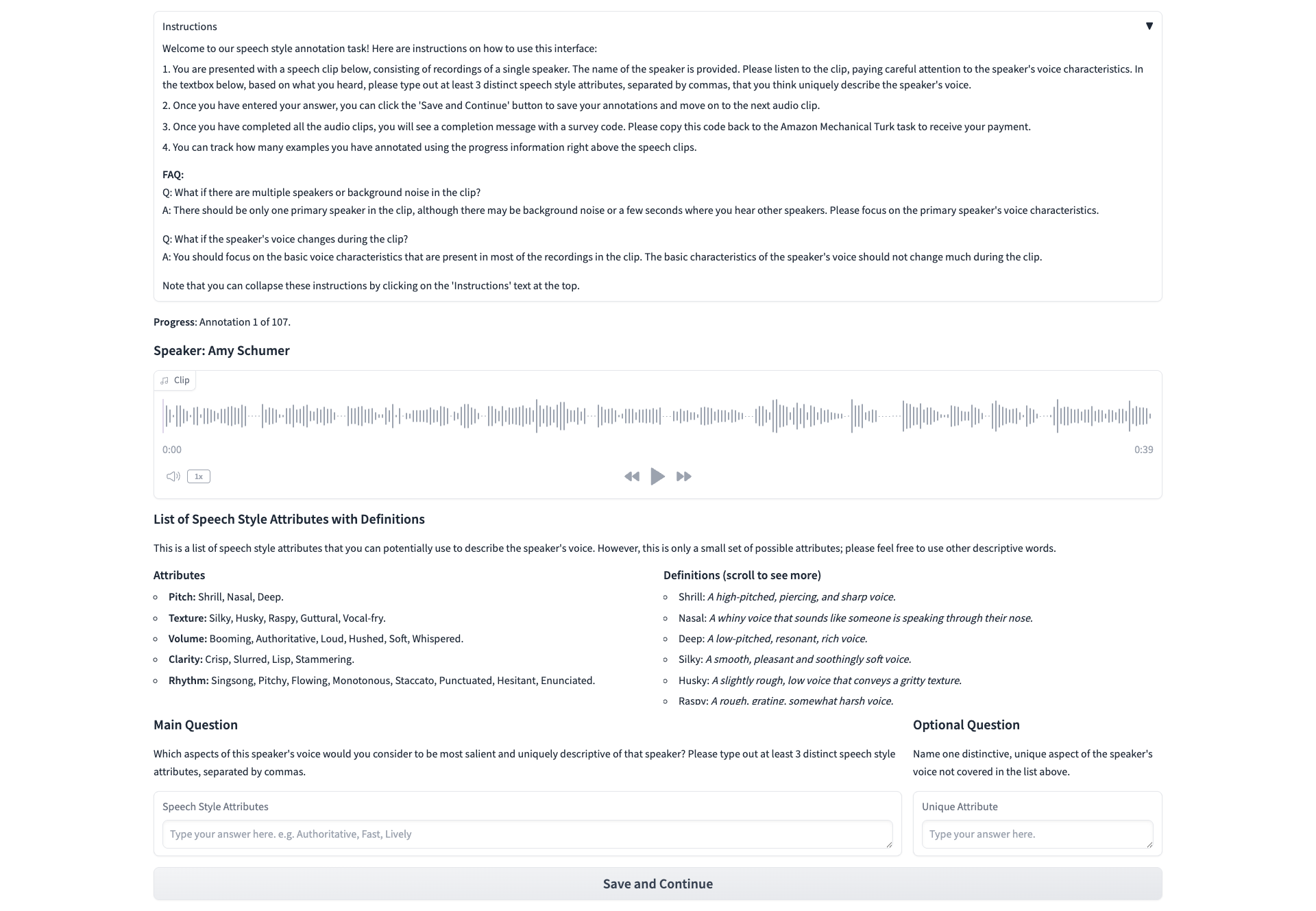}
    \caption{Annotation UI for crowdsourcing abstract intrinsic style tag annotations.}
    \label{fig:intrinsic_annotation_UI}
\end{figure*}

\begin{figure*}[htp!]
    \centering
    \includegraphics[width=\linewidth]{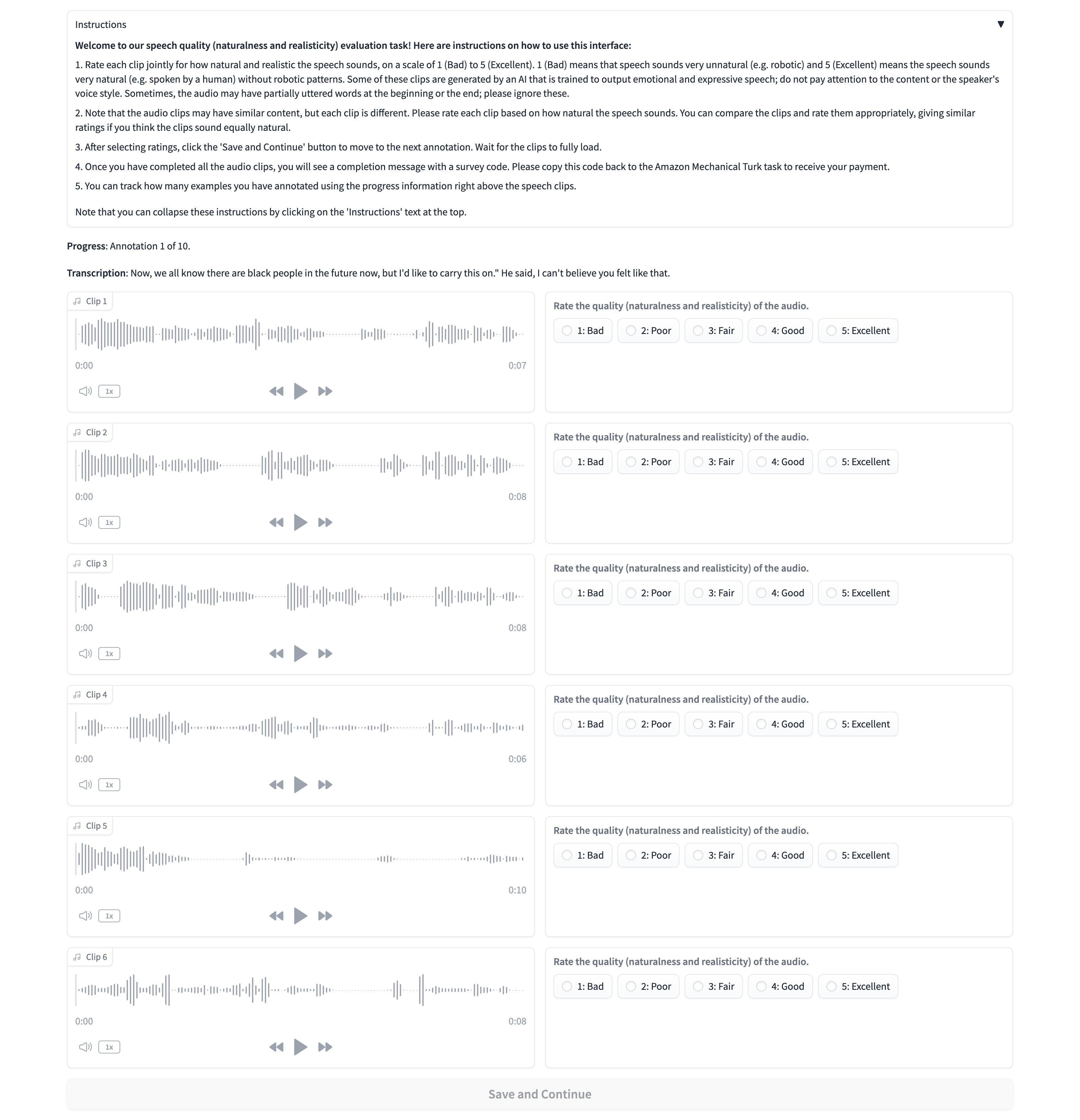}
    \caption{Annotation UI for collecting Naturalness Mean Opinion Score ratings.}
    \label{fig:nmos_evaluation_UI}
\end{figure*}

\begin{figure*}[htp!]
    \centering
    \includegraphics[width=\linewidth]{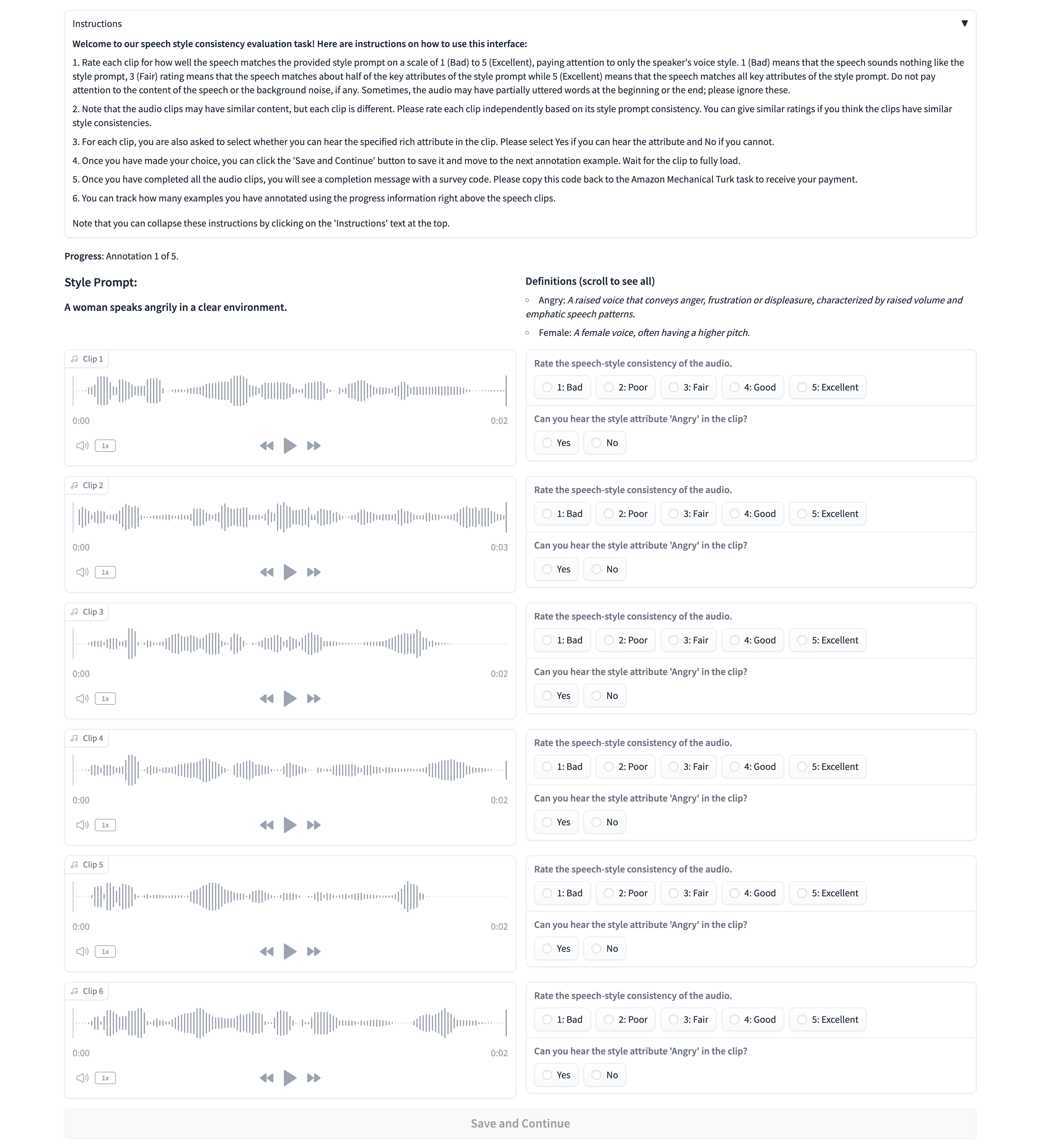}
    \caption{Annotation UI for collecting Consistency Mean Opinion Score and Tag Recall ratings.}
    \label{fig:consistency_mos_UI}
\end{figure*}

\begin{figure*}[htp!]
    \centering
    \includegraphics[width=\linewidth]{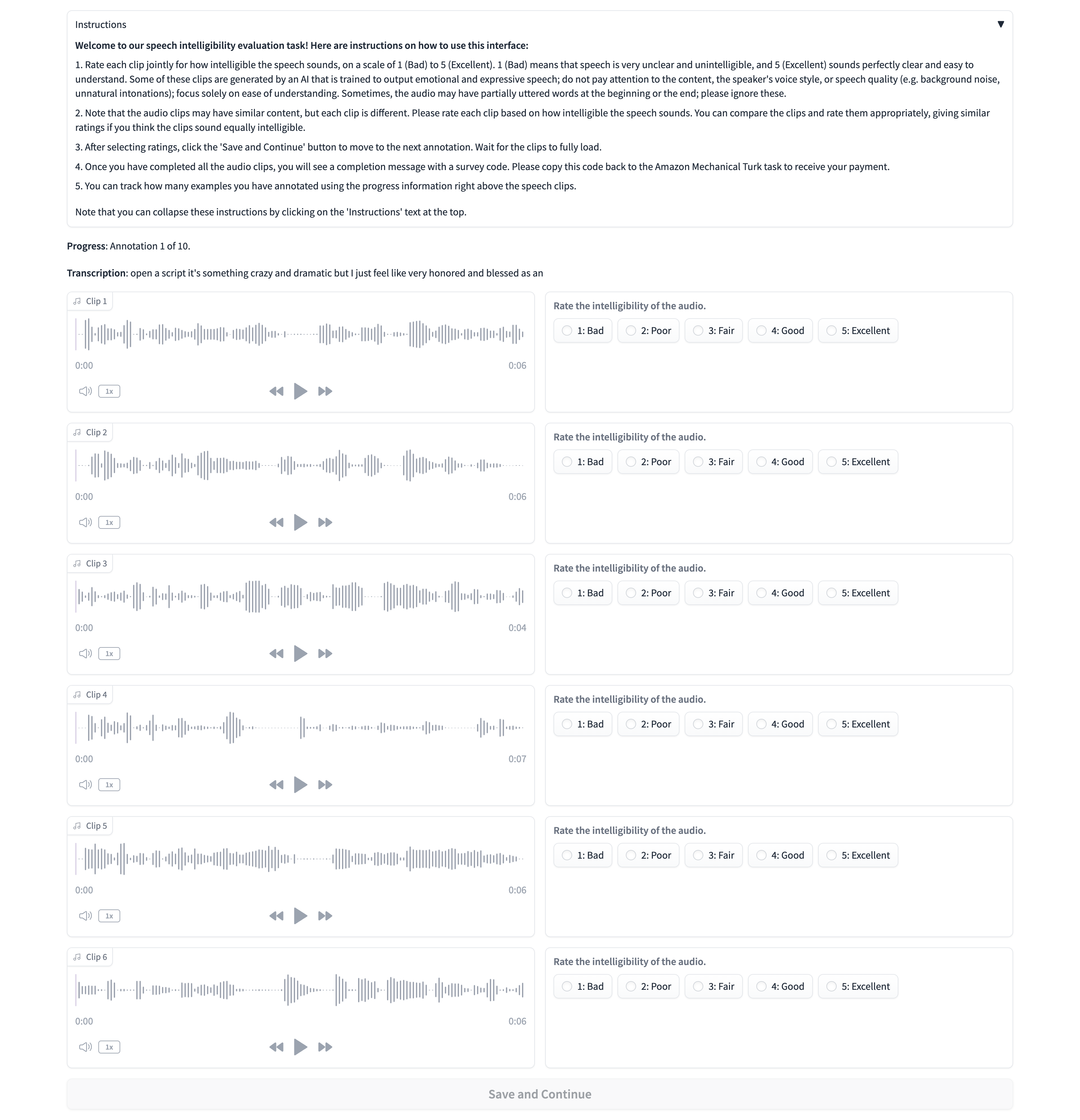}
    \caption{Annotation UI for collecting Intelligibility Mean Opinion Score ratings.}
    \label{fig:imos_evaluation_UI}
\end{figure*}

We display the annotation UIs for qualification task in Figure~\ref{fig:qualification_task_UI}, crowdsourcing abstract intrinsic style tag annotations in Figure~\ref{fig:intrinsic_annotation_UI}, speech quality evaluation in Figure~\ref{fig:nmos_evaluation_UI}, and speech-style consistency evaluation in Figure~\ref{fig:consistency_mos_UI}, and intelligibility evaluation in Figure~\ref{fig:imos_evaluation_UI}.

\section{Dataset Preprocessing}
\label{app:preprocessing}
For all datasets, we filter for audios between $2-30$ seconds.  For data sourced from VoxCeleb, EARS and Expresso, we apply loudness normalization using SoX and PyDub~\footnote{\url{https://sourceforge.net/projects/sox/}, \url{https://github.com/jiaaro/pydub}} such that the peak volume of each audio is $-0.1$ dB. We synthesize transcripts using the Whisper~\citep{radford2022robustspeechrecognitionlargescale} \texttt{large-v3} ASR model for utterances that do not have ground truth transcripts,  We describe dataset-specific preprocessing below:
\subsection{VoxCeleb}
We combine the VoxCeleb1 and VoxCeleb2 datasets. We apply a noise removal model, Voicefixer~\citep{liu2021voicefixer} to all audios, since we observed that a significant proportion of VoxCeleb data is noisy (the median SNR for VoxCeleb data is $31.76$ dB computed by Brouhaha~\citep{lavechin2023brouhaha}; compare  to $59.49, 50.42$ and $61.70$ for Expresso, EARS and LibriTTS-R respectively). We then run a language identification model Lingua~\footnote{\url{https://github.com/pemistahl/lingua-py}} over the transcripts and only keep those examples whose transcripts are identified as English text and discard celebrities with fewer than $10$ English audio clips.

\subsection{Expresso and EARS}
The Expresso and EARS dataset consists of a total of 111 speakers enacting various speaking styles. We discard the \textit{default, narration, non-verbal, interjection} and \textit{vegatative} speaking styles, as they do not possess the styles we are interested in. Some Expresso data is in the form of long dual-channel conversations between two voice actors, which we splice into chunks using Voice Activity Detection metadata provided by the dataset. We discard long freeform EARS examples since they are not labelled with speaking styles. We then remap each speaking style to our tag vocabulary as depicted in Table~\ref{tab:termremapping}.

\subsection{Basic Tagging Thresholds}
\label{sec:thresholds}
\paragraph{\textbf{Pitch:}} low-pitched (male: $<$ 115.7 Hz, female: $<$ 141.6 Hz), high-pitched (male: $>$ 149.7 Hz, female $>$ 184.5 Hz), otherwise medium-pitched.

\paragraph{\textbf{Speed:}} slow: $<$ 11.5 PPS, fast: $>$ 19.1 PPS, otherwise measured.

\paragraph{\textbf{Noise Levels:}} 17.1 dB, 25.4 dB, 33.7 dB, 42.0 dB, 50.2 dB, 58.5 dB, 66.8 dB, 75.0 dB.

\subsection{Scaling Situational Rich Tagging: Details}
\label{sec:dvadirections}
We use emotion-specific dominance-valence-arousal threshold directions in the Expressivity Filtering step and remove transcripts with certain emotion-specific keywords in the Semantic Matching step. These threshold directions and keywords can be found in Table~\ref{tab:emotiondetails}.

\begin{table}[h!]
\centering
\scriptsize
\setlength{\tabcolsep}{4pt} %
\begin{tabular}{ll ll}
\toprule
\textbf{Original} & \textbf{Mapped} & \textbf{Original} & \textbf{Mapped} \\
\midrule
feminine & female & halting & stammering \\
tensed & anxious & relaxed & calm \\
powerful & authoritative & muffled & slurred \\
masculine & male & fluent & flowing \\
weak & hushed & sharp & crisp \\
reassuring & sympathetic & lively & enthusiastic \\
cool & calm & happy & happy, animated \\
laughing & laughing, animated & sad & saddened \\
whisper & whispered & singing & singsong \\
angry & angry, animated & awe & awed \\
bored & bored, passive & desire & desirous, animated \\
projected & loud & fearful & scared \\
amusement & happy & distress & anxious, scared \\
disappointment & saddened, passive & realization & awed \\
amazement & awed & disgust & disgusted \\
fear & scared & anger & angry \\
adoration & admiring & confusion & confused \\
desire & desirous & interest & enthusiastic \\
serenity & calm & contentment & calm, passive \\
sadness & saddened & extasy & happy \\
pain & pained & cuteness & happy \\
relief & calm, passive & pride & admiring \\
embarrassment & anxious & loud & loud \\
\bottomrule
\end{tabular}
\caption{Terms in existing datasets remapped to terms in our vocabulary.}
\label{tab:termremapping}
\end{table}

\begin{table}[h!]
\scriptsize
    \centering
    \setlength{\tabcolsep}{2pt} %
    \begin{tabular}{lccc}
        \toprule
        \textbf{Emotion} & \textbf{A/D} & \textbf{V} & \textbf{Keywords} \\
        \midrule
        Enthusiastic & High & High & enthusiast, excite, eager, energetic, passion \\
        Happy & High & High & happ, joy, cheer, delight, bliss, happy \\
        Angry & High & Low & ang, rage, fury, irritat, frustrat \\
        Saddened & Low & Low & sad, grief, sorrow, mourn, heartbreak \\
        Awed & -- & High & awe, wonder, amaz, astonish, marvel \\
        Calm & Low & -- & calm, peace, seren, relax, tranquil \\
        Anxious & -- & Low & anxi, nerv, uneas, worr, restless \\
        Disgusted & -- & Low & disgus, revolt, repuls, nausea, offend \\
        Scared & High & Low & scar, fear, terror, fright, panick \\
        Confused & -- & -- & confu, bewild, perplex, puzzle, unclear \\
        Bored & Low & -- & bore, dull, uninterest, monoton, tiresom \\
        Sleepy & Low & -- & sleep, drows, fatigu, letharg, slugg \\
        Pained & -- & Low & pain, ache, hurt, agon, torment \\
        Guilt & -- & Low & guilt, blame, shame, remors, regret \\
        Sarcastic & -- & -- & sarca, mock, snark, irony, ridicul \\
        Sympathetic & -- & High & sympath, compass, kind, empath, understand \\
        Admiring & High & High & admir, prais, adore, respect, esteem \\
        Desirous & High & High & desir, crave, long, want, yearn \\
        \bottomrule
    \end{tabular}
    \caption{Mapping of Emotions to Arousal/Dominance and Valence thresholds, along with keywords that are filtered out. Dashes (--) indicate we do not apply a threshold direction.}
    \label{tab:emotiondetails}
\end{table}

\section{Dataset Statistics}
\begin{figure}
    \centering
    \includegraphics[width=0.8\linewidth]{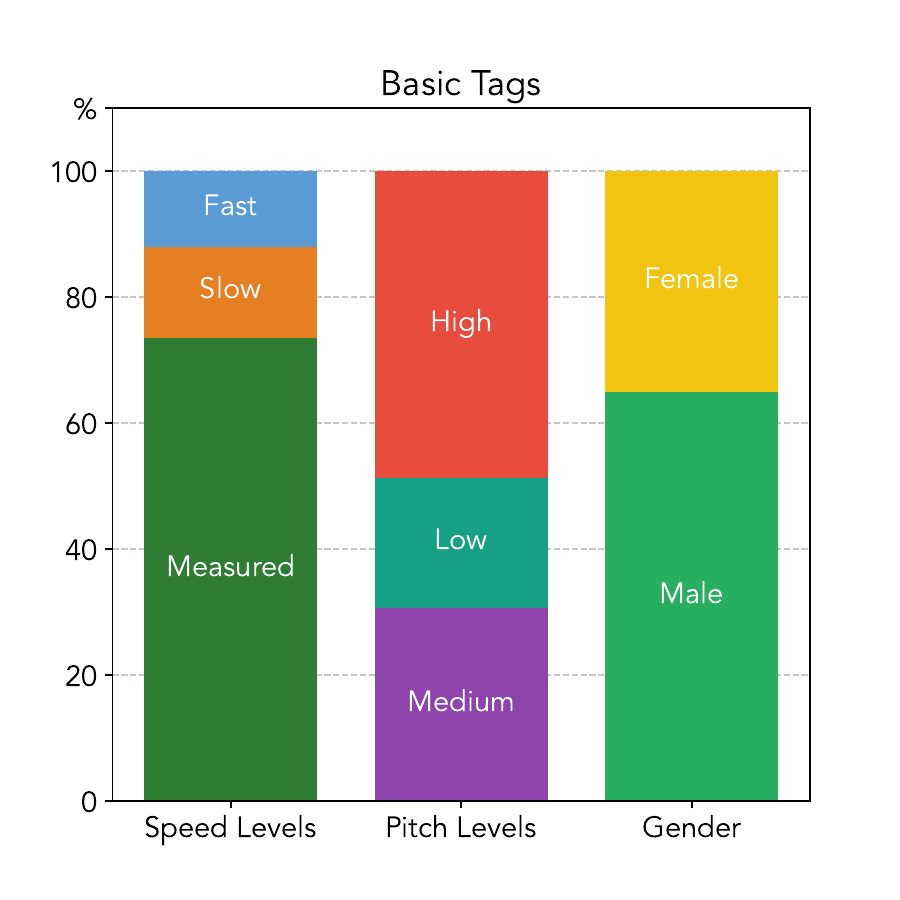}
    \caption{Basic tag distribution in ParaSpeechCaps.}
    \label{fig:basictagstatistics}
\end{figure}

Distributional statistics for basic tags in ParaSpeechCaps is presented in Figure~\ref{fig:basictagstatistics}.

\section{LLM Prompting}
\label{app:promptgeneration}
\subsection{Imperfectly labelling celebrities with style tags}
We use the \texttt{gpt-4-0125-preview} version of GPT-4 via the OpenAI API with the default hyperparameters (temperature $1.0$, top-p $1.0$, maximum $2048$ tokens). We instruct it to output a list of style tags associated with the celebrity's voice with the following prompt, parameterized by \texttt{name}, the name of the celebrity:
\begin{lstlisting}[basicstyle=\scriptsize\ttfamily, breaklines]
Given the name of a famous celebrity or actor, you must retrieve your knowledge about that celebrity's voice and map the voice to a subset of speech style attribute labels provided to you. Here is the list of speech style attribute types you should pay attention to, along with attribute labels for each type:
<attributes>
- **Pitch:** Shrill, Nasal, Deep. 
- **Texture:** Silky, Husky, Raspy, Guttural, Vocal-fry.
- **Volume:** Booming, Authoritative, Loud, Hushed, Soft.
- **Clarity:** Crisp, Slurred, Lisp, Stammering.
- **Rhythm:** Singsong, Pitchy, Flowing, Monotonous, Staccato, Punctuated, Enunciated, Hesitant.
</attributes>

Your task is to associate the celebrity with a subset of these attributes, taking into account how the celebrity always sounds like. Only use the attributes that are extremely salient to the celebrity's voice i.e. their unique speech styles. Don't create any new attributes because you will fail the task if you do so.

The celebrity is {name}. First generate a paragraph of around 5 sentences, within <description> tags, using your knowledge, that describes the salient attributes of {name}'s voice. Then, within <attribute> tags, generate a list of comma-separated speech style attributes, from the above attributes list, that saliently apply to {name}. Use the following format:
<description>
(Description goes here)
</description>
<attribute>
(Comma-separated list of attributes)
</attribute>
\end{lstlisting}

\subsection{Acoustic Matching}
We use the \texttt{gemini-1.5-flash-002} version of Gemini 1.5 Flash via Vertex AI with temperature $1.0$, top-p $0.95$, maximum $2048$ tokens. We instruct it to output its analysis and a rating on a $5$-point Likert scale with a two-part request consisting of the speech clip and the following prompt, parametrized by \texttt{emotion}, the emotion we are querying about:
\begin{lstlisting}[basicstyle=\scriptsize\ttfamily, breaklines]
Analyze the provided speech clip to evaluate how effectively it conveys the emotion {emotion}, focusing on tone of voice and delivery rather than the spoken content.

Key Instructions:
- Focus on Tone: Analyze pitch, tempo, loudness, intonation, and rhythm to judge emotional expression.
- Strength of Emotion: Rate how strongly the tone conveys the emotion on a scale of 1 to 5 (1 = not at all, 5 = very strongly).
- Ignore Content Bias: Evaluate tone and delivery only, disregarding the meaning of the spoken words.

Aspects to Consider:
- Does the pitch and intonation match the energy level of the emotion?
- Is the tempo, rhythm, and loudness appropriate for the emotion?
- Are the tone and delivery consistent with typical characteristics of the emotion?

In your output, start by describing the tone and manner of speaking in the clip. Then, analyze how well the tone aligns with the provided emotion. Finally, rate how strongly the emotion is conveyed on a scale of 1 to 5. To make it easier to parse, format your final answer as follows: "Rating: X/5", where X is the number of your choice.
\end{lstlisting}

\subsection{Extracting Gender and Accent}
We use the \texttt{gpt-4-0125-preview} version of GPT-4 via the OpenAI API with the default hyperparameters (temperature $1.0$, top-p $1.0$, maximum $2048$ tokens). We instruct it to output the celebrity's gender and accent with the following prompt, parameterized by \texttt{name}, the name of the celebrity:
\begin{lstlisting}[basicstyle=\scriptsize\ttfamily, breaklines]
Tell me the accent and the gender of {name} formatted as
Accent: <accent>
Gender: <gender>
\end{lstlisting}

\subsection{Generating Style Prompts}
We use the Mistral-7B-Instruct-v0.2 LLM~\citep{jiang2023mistral7b} to generate prompts via the Dataspeech library with a per-device batch size of $32$ and sample with a temperature of $0.6$, a top-p of $1.0$ with a maximum $256$ new tokens. We instruct the model to generate a style prompt with the following prompt, parametrized by \texttt{all\_tags\_str}, a comma-separated list of style tags:
\begin{lstlisting}[basicstyle=\scriptsize\ttfamily, breaklines]
An audio sample of a person's speech can be described in several ways using descriptive keywords. These keywords may include demographic data about the person (e.g. gender, name, accent) and voice characteristics (e.g. related to pitch, gender, texture and rhythm, volume, clarity, speaking rate, emotion, expressiveness).

You will be provided several keywords that describe the speech sample. Your task is to create a simple text description using the provided keywords that accurately describes the speech sample. Ensure that the description remains grammatically correct, easy to understand, and concise. You can rearrange the keyword order as necessary, and substitute synonymous terms where appropriate. After you are provided the keywords, generate only the description and do not output anything else.  

An example is provided below.
female, confused, hesitant, slightly noisy environment

Description: A woman's speech sounds confused and hesitant, recorded in a slightly noisy environment.

Now, generate a description for the following example:
{all_tags_str}

Description: 
\end{lstlisting}

\section{Discussion Results}
Table~\ref{tab:cfgresults} presents ablation results comparing consistency MOS, Intrinsic and Situational Tag Recalls with and without inference-time classifier-free guidance.

\begin{figure*}
    \centering
    \includegraphics[width=\linewidth]{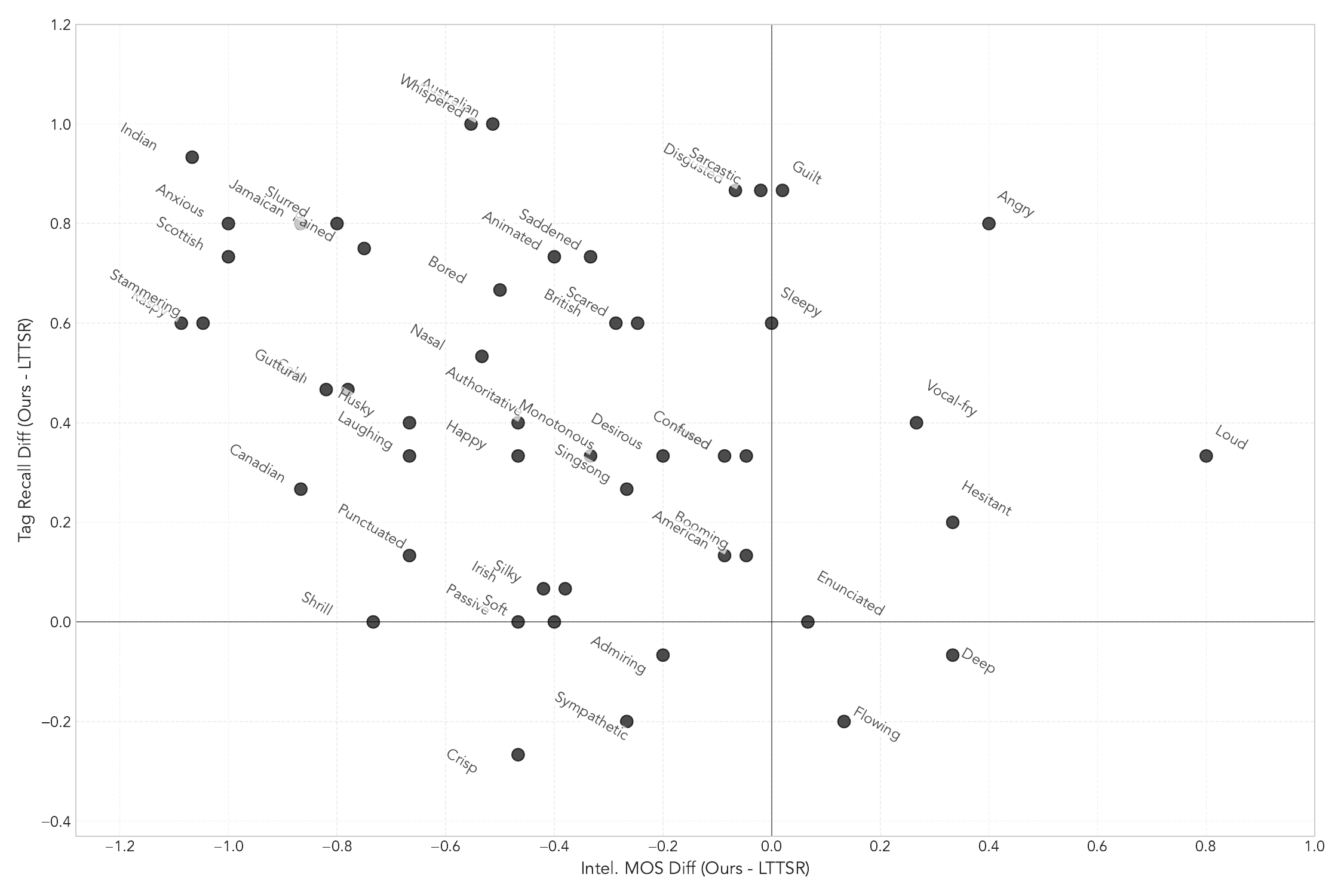}
    \caption{Results showing the difference in the Intelligibility MOS obtained by our Scaled model and the +LTTSR baseline, as well as the difference in the Tag Recall, broken down by tag.}
    \label{fig:intelligibilityvsrecall}
\end{figure*}

Figure~\ref{fig:intelligibilityvsrecall} shows the difference in the Intelligibility MOS obtained by our Scaled model and the +LTTSR baseline, as well as the difference in the Tag Recall, broken down by tag.


\end{document}